\documentclass[11pt,a4paper]{article} 
\usepackage{jheppub}

\usepackage{epsfig,multicol,bbm}
\usepackage{graphicx}
\usepackage{amsmath}
\usepackage{bm}
\usepackage{multirow}
\allowdisplaybreaks[1]

\newcommand\fverb{\setbox\fverbbox=\hbox\bgroup\verb}
\newcommand\fverbdo{\egroup\medskip\noindent%
			\fbox{\unhbox\fverbbox}\ }
\newcommand\fverbit{\egroup\item[\fbox{\unhbox\fverbbox}]}
\newbox\fverbbox

\title{Inclusive double-quarkonium production
at the Large Hadron Collider
}

\author[a]{P. Ko,}
\author[b]{Jungil Lee,}
\author[a]{and Chaehyun Yu}

\affiliation[a]{School of Physics, KIAS, \\ Seoul 130-722, Korea}
\affiliation[b]{Department of Physics, Korea University, \\ 
Seoul 136-701, Korea}

\emailAdd{pko@kias.re.kr}
\emailAdd{jungil@korea.ac.kr}
\emailAdd{chyu@kias.re.kr}

\arxivnumber{1007.3095}	

\abstract{
Based on the nonrelativistic QCD (NRQCD) factorization formalism,
we investigate inclusive productions of two spin-triplet $S$-wave quarkonia
$pp\to 2J/\psi+X$, $2\Upsilon+X$, and $J/\psi+\Upsilon+X$ at the CERN
Large Hadron Collider. The total production rates integrated over the
rapidity ($y$) and transverse-momentum ($p_T$) ranges
$|y|<2.4$ and $p_T<50$\,GeV are
predicted to be 
$\sigma[pp\to 2J/\psi+X]=$         22 (35)\,nb,
$\sigma[pp\to 2\Upsilon+X]=$       24 (49)\,pb, and
$\sigma[pp\to J/\psi+\Upsilon+X]=$  7 (13)\,pb
at the center-of-momentum energy $\sqrt{s}=7$ (14)\,TeV. In order to provide
predictions that can be useful in both small- and large-$p_T$ regions, 
we do not employ the fragmentation approximation and we include
the spin-triplet $S$-wave color-singlet and  color-octet channels
for each quarkonium final state at leading order in the strong coupling.
The $p_T$ distributions of $pp\to 2J/\psi+X$ and $2\Upsilon+X$
in the low-$p_T$ region are dominated by the color-singlet contributions.
At leading order in the strong coupling, the color-singlet channel is
absent for $pp\to J/\psi+\Upsilon+X$. Therefore, the process
$pp\to J/\psi+\Upsilon+X$ may provide a useful probe to the color-octet
mechanism of NRQCD.}
\begin{document} 
\maketitle
\section{Introduction\label{sec:intro}}
Understanding the heavy-quarkonium production mechanism has been
a longstanding problem in QCD, albeit considerable efforts both in theory
and experiments. The color-singlet model~\cite{%
Einhorn:1975ua,Ellis:1976fj,Chang:1979nn,Berger:1980ni,Baier:1981zz}
is most intuitive,
and was the first attempt to describe the data in early days in QCD. 
This has been implemented as a leading-order (LO) contribution
in the nonrelativistic QCD (NRQCD) factorization approach~\cite{BBL}
as $v_Q\to 0$, where $v_Q$ is half the relative velocity of the
heavy quark $Q$ or heavy antiquark $\bar{Q}$ in the 
rest frame of the quarkonium $H$. Another phenomenological
approach is the color-evaporation model~\cite{CEM,Barger:1979js}.
Although it is difficult to make the color-evaporation model
systematic within QCD, this approach may capture an important point
in the heavy-quarkonium production. This is especially the case, because
of the general failure of local duality between quarks/gluons and hadrons.

A salient feature of NRQCD is that color-octet contributions can be
included in a systematic fashion. The velocity-scaling rules (VSR)
of NRQCD scale the numerical value for a long-distance NRQCD matrix
element $\langle O^{H}_n(^{2s+1}L_J)\rangle$, which accounts for the transition
probability of the heavy-quark-antiquark pair $Q\bar{Q}_n({}^{2s+1}L_J)$
to evolve into the quarkonium $H$ in the asymptotic future, in powers of $v_Q$.
Here, $s$, $L$, and $J$ are the quantum numbers for the spin, the orbital
angular momentum, and the total angular momentum of the $Q\bar{Q}$ pair,
respectively, and $n=1$ (8) stands for the color-singlet (color-octet) state.
In combination with the power counting in the strong coupling constant
$\alpha_s$, physical observables can be expanded in double power series 
in $\alpha_s$ and $v_Q$. In particular, in NRQCD the infrared (IR)
divergence in the $P$-wave quarkonium $\chi_{Q_J}$ decay in the
color-singlet model is absorbed by renormalizing the color-octet
matrix element $\langle O^{\chi_{Q_J}}_8(^{3}S_1)\rangle$, resulting
in the decay rate free of an IR divergence~\cite{Bodwin:1992ye}.

The color-octet mechanism of NRQCD also enabled one to explain the
large discrepancy between the data and the color-singlet-model prediction
for prompt spin-triplet $S$-wave charmonium production at the Fermilab
Tevatron \cite{Braaten:1994vv} at large transverse momentum ($p_T$). 
It was possible by allowing a $c\bar{c}_8(^{3}S_1)$ pair, that was
created from the fragmentation of a virtual gluon, to make a transition
into the spin-triplet $S$-wave quarkonium $H$. The corresponding matrix
element $\langle O^{H}_8(^{3}S_1)\rangle$ was determined from the
large-$p_T$ data. For the lower-$p_T$ region, other color-octet
contributions  $c\bar{c}_8(^{1}S_0)$ and $c\bar{c}_8(^{3}P_J)$ as well
as the color-singlet $[c\bar{c}_1(^{3}S_1)]$ channel may also
contribute and a linear combination $M_r$ of the two matrix elements 
$\langle O^{H}_8(^{1}S_0)\rangle$ and $\langle O^{H}_8(^{3}P_J)\rangle$
was fit to the Tevatron data. The resolution of the large-surplus
puzzle of prompt $J/\psi$ at large $p_T$ by using the gluon fragmentation
into $c\bar{c}_8(^{3}S_1)$ was followed by the prediction that the
polarization of prompt $J/\psi$ must be transversely polarized 
\cite{Braaten:1999qk}, which has not been confirmed by the
data~\cite{Affolder:2000nn,Abulencia:2007us}.

However, the values for the color-octet matrix elements 
$\langle O^{H}_8(^{3}S_1)\rangle$, $\langle O^{H}_8(^{1}S_0)\rangle$,
and $\langle O^{H}_8(^{3}P_J)\rangle$ are known not quite accurately
\cite{review} and there are several indications that these three matrix
elements fit to the Tevatron data have been overestimated. Considerable
higher-order corrections in $\alpha_s$ to the color-singlet contribution to
the prompt-$J/\psi$ production rate at the Tevatron have been reported
although there is still a sizable room for the color-octet contribution,
especially at large $p_T$ \cite{Campbell:2007ws,Gong:2008sn}.
The nonobservation of strongly transverse polarization of prompt
$J/\psi$ also indicates that the value for the matrix element
$\langle O^{H}_8(^{3}S_1)\rangle$ has been overestimated. This puzzle
remains unresolved in spite of recent theoretical improvements 
\cite{Gong:2008sn,Gong:2008ft}. An analysis at next-to-leading-order 
(NLO) in $\alpha_s$ of the inclusive-$J/\psi$ cross section at the $B$
factories~\cite{Ma:2008gq} showed that the color-singlet contribution agrees
with the data measured by the Belle Collaboration~\cite{:2009nj}
within uncertainties.
According to this analysis, the upper bound of the color-octet matrix
elements $\langle O_8^{J/\psi}(^1S_0)\rangle$ or 
$\langle O_8^{J/\psi}(^3P_0)\rangle$ is much smaller than those determined by
other experiments~\cite{Zhang:2009ym}.
Recent analyses of inclusive $J/\psi$ photoproduction
at NLO accuracies in $\alpha_s$ show that the color-singlet contribution fails
to describe various features of the data at HERA~\cite{Artoisenet:2009xh}
and that the color-octet mechanism explains the H1 data in spite of the
poor knowledge of the color-octet matrix elements~\cite{Butenschoen:2009zy}.
In summary, there are no conclusive constraints on the color-octet matrix
elements, and it is still controversial if the color-octet mechanism
makes a substantial contribution to the inclusive production of a
quarkonium. It would be highly desirable to identify some processes which
depend only on a very few color-octet matrix elements so that one can
derive strong phenomenological constraints.

Since the observation of exclusive double-quarkonium final states
by the Belle Collaboration~\cite{Abe:2002rb}, production of
double-quarkonium states
in $e^+e^-$ collisions has lead remarkable progress in understanding
the interplay of relativistic corrections and NLO corrections in
$\alpha_s$~\cite{double1,double2,lightcone,doubleqcd,doublerel}.
The study has recently been extended to hadroproduction like the inclusive
productions of double $J/\psi$'s, double $\Upsilon$'s\footnote{Throughout this
paper we suppress the identifier $(1S)$ for $\Upsilon(1S)$.},
and a $B_c^{(*)} \bar{B}_c^{(*)}$ pair at the Tevatron and the
CERN Large Hadron Collider (LHC)~\cite{Qiao:2002rh,Li:2009ug,Qiao:2009kg}. 
For the
double-$J/\psi$ production at LO in $\alpha_s$, the color-singlet contribution
dominates over the color-octet one at $p_T\lesssim$ 8\,GeV~\cite{Li:2009ug}.
On the contrary, in the case of the double-$\Upsilon$ production, the
color-octet contribution dominates over the whole range of
$p_T$~\cite{Li:2009ug}.
These predictions may be useful to study the color-octet
mechanism in quarkonium production. In refs.~\cite{Li:2009ug,Qiao:2009kg},
the authors employed the gluon-fragmentation approximation 
in order to estimate the color-octet contribution.
Although the approximation may give a reliable prediction at large $p_T$, 
it should lose its predictive power at low values of $p_T$.

In this work, we study inclusive productions of spin-triplet $S$-wave
heavy-quarkonium pairs at the LHC in $pp$ collisions at the
center-of-momentum (CM) energies $\sqrt{s}=$ 7 and 14\,TeV. The final
states considered are double $J/\psi$'s, double $\Upsilon$'s,
and $J/\psi+\Upsilon$.
The analysis is carried out at LO in $v_Q$ and in $\alpha_s$
without employing the gluon-fragmentation approximation.
This may allow us to provide a more reliable predictions for
the production rates in the intermediate $p_T$ region, where
the production rate is large in comparison with the large-$p_T$ region.
The $J/\psi + \Upsilon$ final state has a special feature that the
color-singlet contribution is absent at LO in $\alpha_s$.
Hence this process might be a clean probe to color-octet mechanism.
Conversely, if $J/\psi + \Upsilon$ events are not observed at the
proposed level, it may lower the current upper bounds for the
color-octet matrix elements significantly.

This paper is organized as follows. In section~\ref{sec:strategy}, we describe
basic strategies to compute the inclusive cross sections for $pp\to 2J/\psi+X$,
$2\Upsilon+X$, and $J/\psi+\Upsilon+X$. We list various input parameters in
section~\ref{sec:numerical} and our predictions are given in 
section~\ref{sec:predictions}. We conclude in section~\ref{sec:discussion} and
provide relevant parton-level cross-section formulas in the appendix.
\section{Inclusive double-quarkonium production\label{sec:strategy}}
The NRQCD factorization formula for the differential cross section $d \sigma$
of inclusive double-quarkonium production in proton-proton collisions,
$pp\to H_1(p_1)+H_2(p_2)+X$, has the following schematic form:
\begin{equation}
d \sigma[pp\to H_1+H_2+X]
  =
\sum_{a,\,b,\,n_1,\,n_2}
f_{a/p}\otimes f_{b/p}\otimes
d \hat{\sigma}[ab \to \mathcal{Q}_1^{n_1}+\mathcal{Q}_2^{n_2}]
\,
\langle O^{H_1}_{n_1} \rangle
\langle O^{H_2}_{n_2} \rangle,
\label{xsec}%
\end{equation}
where $p_i$ is the momentum of the quarkonium $H_i$, $f_{a/p}(x_a,\mu)$ is
the parton distribution function (PDF) for the parton $a$ 
with the longitudinal momentum
fraction $x_a$ with respect to the proton, $\mu$ is the factorization scale, 
the symbol $\otimes$ indicates the convolution over the partons' longitudinal
momentum fractions $x_a$ and $x_b$, and the summation is over all possible
combinations of partons $a$ and $b$. 
The parton-level differential cross section 
$d \hat{\sigma}[{ab}\to \mathcal{Q}_1^{n_1}+\mathcal{Q}_2^{n_2}]$ is the
short-distance coefficient, which is perturbatively calculable in powers of
$\alpha_s$. Here, $\mathcal{Q}_i^{n_i}\equiv Q\bar{Q}_{n}({}^{2s+1}L_J)$
denotes the $i$-th $Q\bar{Q}$ pair with the momentum $p_i$ and with the
spectroscopic index $n_i$, which evolves asymptotically into $H_i$.  
The dependence of the long-distance nature of the heavy quarkonium $H_i$ is
factored into the NRQCD matrix element $\langle O_{n_i}^{H_i} \rangle$. 
The expression (\ref{xsec}) is a power series in $\alpha_s$, $v_c$, and  $v_b$.

At LO in $\alpha_s$, only $g g$ fusion and $q \bar{q}$ annihilation contribute
to the parton processes $ab\to \mathcal{Q}_1^{n_1}+\mathcal{Q}_2^{n_2}$ if both
$\mathcal{Q}_1$ and $\mathcal{Q}_2$ are of the same flavor. The 
processes with the initial partons $ab,\,ba=gq$ and $g\bar{q}$ are missing
at this order. At the CM energies $\sqrt{s}=7\,$ and $14$\,TeV of
the LHC, one probes the small-$x$ region of the PDF, where the gluon
contribution dominates over the quark contents. Therefore, we ignore the parton
processes initiated from $q \bar{q}$, $gq$, and $g\bar{q}$ states and consider
only the $gg$ initial states in this work.

\begin{figure}
\begin{center}
\epsfig{file=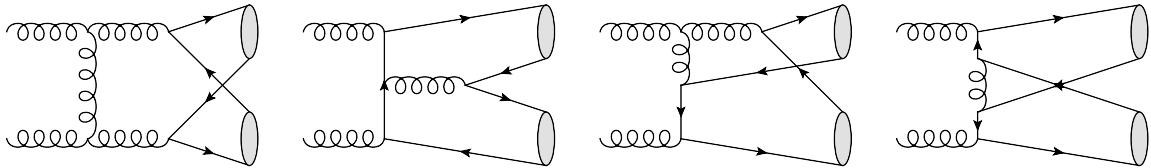,width=\textwidth}
\caption{\label{fignonfrag}%
Typical Feynman diagrams for the nonfragmentation 
contribution to $pp\to 2H+X$ at LO in $\alpha_s$. 
Only gluon-gluon fusion diagrams are shown. The quark lines represent 
the charm (bottom) quark for $H=J/\psi$ $(\Upsilon)$.}
\end{center}
\end{figure}

\begin{figure}
\begin{center}
\epsfig{file=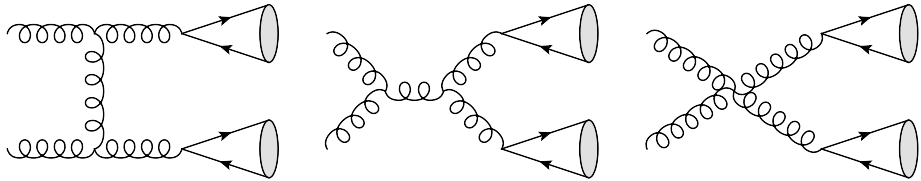,width=0.8\textwidth}
\caption{\label{figfrag}%
Typical Feynman diagrams for the gluon-fragmentation contribution to
double-quarkonium production processes
$pp\to 2J/\psi+X$, $2\Upsilon+X$, and $J/\psi+\Upsilon+X$
at LO in $\alpha_s$.}
\end{center}
\end{figure}
\subsection{Inclusive identical spin-triplet $S$-wave quarkonium pair production%
\label{sec:ii-identical}}
Let us first consider inclusive identical spin-triplet $S$-wave quarkonium
pair production $pp\to 2H+X$, where $H$ is $J/\psi$ or $\Upsilon$. In this
case, the LO parton process is of order $\alpha_s^4$. As is stated earlier
in this section, we consider only the two-gluon initial states among various
parton processes. The corresponding Feynman diagrams are classified into two
groups: one is the nonfragmentation contribution, whose typical diagrams are
shown in figure~\ref{fignonfrag}, and the other is the gluon-fragmentation
contribution shown in figure~\ref{figfrag}. The sum of the two sets of
diagrams can only make a gauge-invariant amplitude. In the limit that the
invariant mass of $\mathcal{Q}_i$ vanishes, the set of fragmentation diagrams
approaches a gauge-invariant subset.

According to VSR,  the leading spectroscopic states of $\mathcal{Q}^{n_1}_1$
and $\mathcal{Q}^{n_2}_2$ in the velocity expansion for $H=J/\psi$ or
$\Upsilon$ production are both $Q\bar{Q}_1({}^3S_1)$ unless there is any
enhancement factor for other spectroscopic states to compensate the power
suppression compared to this color-singlet contribution. For the
nonfragmentation diagrams in figure~\ref{fignonfrag}, the leading contribution
must be the color-singlet channel $Q\bar{Q}_1({}^3S_1)+Q\bar{Q}_1({}^3S_1)$.
In the case of the gluon-fragmentation contribution 
$Q\bar{Q}_8({}^3S_1)+Q\bar{Q}_8({}^3S_1)$ in figure~\ref{figfrag}, the large
kinematic enhancement factor for the color-octet spin-triplet $S$-wave 
[$Q\bar{Q}_8({}^3S_1)$] contribution, which is suppressed by a relative order
$v_Q^8$ compared to the color-singlet contribution, actually overcomes the
suppression factor in the large-$p_T$ region. Here, $Q=c\,(b)$ for
$H=J/\psi\,(\Upsilon)$. However, such enhancement factors do not appear in
the mixed channels $Q\bar{Q}_8({}^3S_1)+Q\bar{Q}_1({}^3S_1)$ and
$Q\bar{Q}_1({}^3S_1)+Q\bar{Q}_8({}^3S_1)$ that are suppressed by $v_Q^4$
compared to the color-singlet channel. There are two other color-octet
contributions, the spin-singlet $S$-wave [$Q\bar{Q}_8({}^1S_0)$] and the
spin-triplet $P$-wave [$Q\bar{Q}_8({}^3P_J)$], where $J=0$, 1, and 2, which
are suppressed by $v_Q^3$ and $v_Q^{4}$, respectively, compared to the
color-singlet contribution. Nevertheless, we ignore these two contributions
because they do not have any large enhancement factor to compete with either 
$Q\bar{Q}_1({}^3S_1)$ or $Q\bar{Q}_8({}^3S_1)$ contribution.

In summary, the contributions to the identical-quarkonium pair production
that we consider in this work are
$Q\bar{Q}_1({}^3S_1)+Q\bar{Q}_1({}^3S_1)$ and
$Q\bar{Q}_8({}^3S_1)+Q\bar{Q}_8({}^3S_1)$.
There are 31 Feynman diagrams that contribute to the color-singlet channel
$gg\to Q\bar{Q}_1({}^3S_1) + Q\bar{Q}_1({}^3S_1)$ whose typical diagrams are
shown in figure~\ref{fignonfrag}. For the color-octet channels
$gg\to Q\bar{Q}_8({}^3S_1)+Q\bar{Q}_8({}^3S_1)$, there are 72 Feynman
diagrams, some of which are shown in figures~\ref{fignonfrag} and \ref{figfrag}.

The differential cross section in eq.~(\ref{xsec}) is applicable to both
$pp\to 2J/\psi+X$ and $pp\to 2\Upsilon+X$ if we substitute $Q=c$ and $b$ in
the short-distance coefficients and $H=J/\psi$ and $\Upsilon$ in the NRQCD
matrix elements, respectively. In the range where $p_T$ is not sufficiently
large, the nonfragmentation contributions in figure~\ref{fignonfrag} are not
suppressed. Therefore, we compute the parton cross sections 
$d \hat{\sigma}[ab \to \mathcal{Q}_1^{n_1}+\mathcal{Q}_2^{n_2}]$
for the relevant channels including all types of diagrams shown in 
figures~\ref{fignonfrag} and \ref{figfrag}. The parton-level
cross sections for the color-singlet contributions are given in
refs.~\cite{Li:2009ug,Qiao:2009kg} and we have reproduced the results
explicitly. The expression for the parton-level cross section for
the color-octet channel $gg\to Q\bar{Q}_8({}^3S_1)+Q\bar{Q}_8({}^3S_1)$
is given in appendix~\ref{app:same}.
In the large-$p_T$ region, the color-octet gluon-fragmentation
diagrams in figure~\ref{figfrag} dominate and they can be computed by employing
the fragmentation approximation of the gluon that fragments into a
$Q\bar{Q}_8({}^3S_1)$ pair. In refs.~\cite{Li:2009ug,Qiao:2009kg}, such an
approximation has been used to compute the cross section for the 
double-quarkonium production at large $p_T$. In these references the parton
cross sections for the real-gluon final states $ab\to gg$ are convolved with
the gluon-fragmentation function. In this work, we do not employ the
fragmentation approximation and compute the complete set of order-$\alpha_s^4$ 
Feynman diagrams for
$Q\bar{Q}_1({}^3S_1)+Q\bar{Q}_1({}^3S_1)$ and
$Q\bar{Q}_8({}^3S_1)+Q\bar{Q}_8({}^3S_1)$ channels.
Therefore, our calculations provide predictions for these processes
which can be compared with $p_T$ spectra in a wider range of $p_T$ including
the small $p_T$ region. Further inclusion of the order-$\alpha_s^4$
color-octet channels that we have ignored in this work may improve the
predictions in the intermediate-$p_T$ region, not modifying low- and
large-$p_T$ spectra significantly. This requires extensive calculations of
formidably many Feynman diagrams and is beyond the scope of this work. 
\begin{figure}
\begin{center}
\epsfig{file=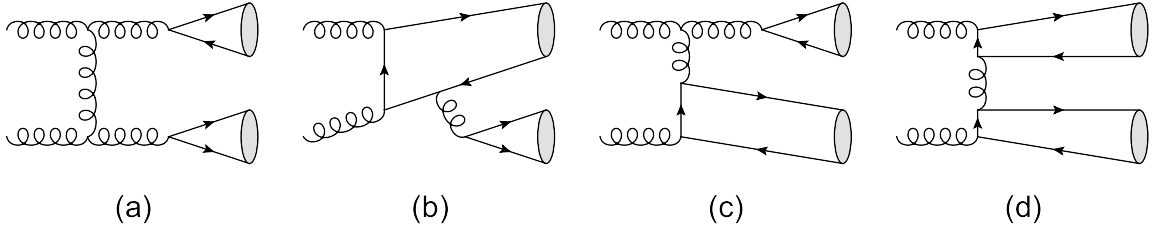,width=\textwidth}
\\[-2.5ex]
\caption{\label{figpu}%
Typical Feynman diagrams contributing to
$pp\to J/\psi+\Upsilon$ at LO in $\alpha_s$.}
\end{center}
\end{figure}
\subsection{$J/\psi + \Upsilon$ production%
\label{sec:ii-different}}
Next we consider the inclusive double-quarkonium production with different
flavors, $pp\to J/\psi + \Upsilon+X$. Like the case of the same flavor,
order-$\alpha_s^4$ diagrams are leading in powers of $\alpha_s$. 
Typical Feynman diagrams for the gluon-initiated parton processes
$gg\to \mathcal{Q}_1^{n_1}+\mathcal{Q}_2^{n_2}$ are given in figure~\ref{figpu},
where $\mathcal{Q}_1^{n_1}$ ($\mathcal{Q}_2^{n_2}$) stands for the $c\bar{c}$
($b\bar{b}$) pair. At this order, the color-singlet channel
$c\bar{c}_1({}^3S_1)+b\bar{b}_1({}^3S_1)$ is absent.

\subsubsection{Color-octet contributions\label{sec:psi-ups-octet}}
We first consider the color-octet contribution
$c\bar{c}_8({}^3S_1)+b\bar{b}_8({}^3S_1)$, whose ``velocity-scaling factor'' 
$\mathcal{V}= v_c^4 v_b^4$. There are 36 Feynman diagrams that contribute
to this channel and we show typical ones in figure~\ref{figpu}. At
large $p_T$ the double-fragmentation contribution in figure~\ref{figpu}\,(a)
dominates the cross section because of the large kinematic enhancement. Here,
double fragmentation denotes that both $c\bar{c}$ and $b\bar{b}$ pairs are
produced via gluon fragmentation. Figure~\ref{figpu}\,(b) also represents mixed
contributions $c\bar{c}_8({}^3S_1)+b\bar{b}_1({}^3S_1)$ and
$c\bar{c}_1({}^3S_1)+b\bar{b}_8({}^3S_1)$, each of which has 6 Feynman diagrams.
Their velocity-scaling factors are $v_c^4$ and $v_b^4$, respectively,
that are enhanced by either $1/v_b^4$ or $1/v_c^4$ compared to
the channel $c\bar{c}_8({}^3S_1)+b\bar{b}_8({}^3S_1)$.
Therefore, if $p_T$ is not large enough, then
these mixed contributions must dominate over the color-octet channel
$c\bar{c}_8({}^3S_1)+b\bar{b}_8({}^3S_1)$ by the enhancement
factors $1/v_c^4$ or $1/v_b^4$, while the double-fragmentation contribution
dominates at large $p_T$.

There are other contributions that are power suppressed compared to the
contributions listed above. Such contributions are
$c\bar{c}_8({}^3S_1)+b\bar{b}_8({}^{2s+1}L_J)$ and
$c\bar{c}_8({}^{2s+1}L_J)+b\bar{b}_8({}^3S_1)$,
where ${}^{2s+1}L_J={}^1S_0$ or ${}^3P_J$. The corresponding Feynman diagrams
representing these contributions are shown in 
figure~\ref{figpu}\,(b), (c), and (d), with $\mathcal{V}=v_c^\alpha v_b^\beta$.
Here, $\alpha, \beta = 3$ for ${}^{2s+1}L_J = {}^1S_0$ and 4 
for ${}^3S_1$ and ${}^3P_J$.
The velocity-scaling factors of these contributions are approximately 
similar to
that of the channel $c\bar{c}_8({}^3S_1)+b\bar{b}_8({}^{3}S_1)$
but they do not have any double-fragmentation contribution. At small values
of $p_T$, these contributions are suppressed at least by either $v_c^3$ or
$v_b^3$ compared to the mixed channels
$c\bar{c}_1({}^3S_1)+b\bar{b}_8({}^{3}S_1)$ and
$c\bar{c}_8({}^3S_1)+b\bar{b}_1({}^{3}S_1)$.
Although the single-fragmentation channel in figures~\ref{figpu}\,(b) and (c)
may grow up
at large $p_T$, that contribution is dominated by the double-fragmentation
[figure~\ref{figpu}\,(a)] by a factor of $(m_c/p_T)^4$ or $(m_b/p_T)^4$. Hence,
it is consistent to ignore $c\bar{c}_8({}^3S_1)+b\bar{b}_8({}^{2s+1}L_J)$ and
$c\bar{c}_8({}^{2s+1}L_J)+b\bar{b}_8({}^3S_1)$ channels over the whole $p_T$
range.

The last color-octet contributions we can consider are the channels 
$c\bar{c}_8({}^{2s+1}L_J)+b\bar{b}_8({}^{2s'+1}L'_{J'})$,
where ${}^{2s+1}L_J$ and ${}^{2s'+1}L'_{J'}$ are ${}^1S_0$ or ${}^3P_J$.
A typical Feynman diagram of these channels is shown in figure~\ref{figpu}\,(d).
These contributions are, again, scaled by $\mathcal{V}=v_c^\alpha v_b^\beta$,
where $\alpha, \beta = 3$ for ${}^{2s+1}L_J = {}^1S_0$ and 4 for ${}^3P_J$.
Thus these contributions are suppressed compared to the color-singlet one.
Because they do not have any fragmentation contributions, they are dominated by
the double-fragmentation [figure~\ref{figpu}\,(a)] by $(m_c/p_T)^4 (m_b/p_T)^4$
at large $p_T$. Therefore, we ignore these channels, too.

In summary, all of the channels that we consider
for the process $p p \to J/\psi+\Upsilon + X$ in this work are
$c\bar{c}_8({}^3S_1)+b\bar{b}_8({}^3S_1)$,
$c\bar{c}_8({}^3S_1)+b\bar{b}_1({}^3S_1)$, and
$c\bar{c}_1({}^3S_1)+b\bar{b}_8({}^3S_1)$.
The parton-level differential cross sections for these processes
are given in appendix~\ref{app:diff}.
\begin{figure}
\begin{center}
\epsfig{file=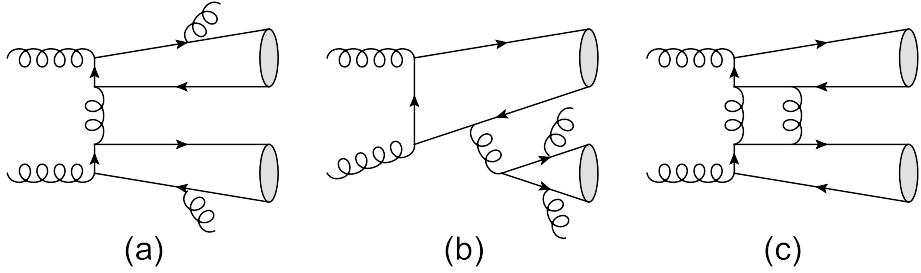,width=0.8\textwidth}
\caption{\label{figjusing}%
Typical Feynman diagrams for 
the color-singlet contribution to
$gg\to J/\psi + \Upsilon +X$ at order $\alpha_s^6$.}
\end{center}
\end{figure}
\subsubsection{Color-singlet contributions}
As shown in figure~\ref{figjusing}, the LO color-singlet contribution to
$pp\to J/\psi+\Upsilon+X$ is of order $\alpha_s^6$ and is suppressed
compared to the color-octet contribution by $\alpha_s^2$.
The corresponding parton processes consist of 3 types: 
Type A, which is shown in figure~\ref{figjusing}\,(a), is
$gg\to c\bar{c}_1({}^3S_1)+b\bar{b}_1({}^3S_1)+gg$, where the initial
gluons are attached to the charm- and bottom-quark lines one by one,
a single virtual gluon connects the two quark lines, and each quark
line emits a single real gluon.  
Type B is, again,
$gg\to c\bar{c}_1({}^3S_1)+b\bar{b}_1({}^3S_1)+gg$, where both initial
gluons are attached to a single quark line, the other quark line emits two
real gluons, and a virtual gluon connects the two quark lines. 
A Feynman diagram for Type B is shown in figure~\ref{figjusing}\,(b)
and the two types A and B interfere at the amplitude level.
Type C, which is shown in figure~\ref{figjusing}\,(c),
 is $gg\to c\bar{c}_1({}^3S_1)+b\bar{b}_1({}^3S_1)$,
where the initial gluons are attached to the charm- and bottom-quark
lines one by one and two virtual gluons connect the two quark lines
simultaneously. 

If we consider the powers in $\alpha_s$ only, then this color-singlet
contribution must be suppressed compared to the color-octet processes listed
in section~\ref{sec:psi-ups-octet}. To assure that this argument is valid, we
can make a rough estimate of the color-singlet contribution to the inclusive
$J/\psi+\Upsilon$ production rate at hadron colliders.

We first classify the scaling behavior of the amplitude for the
double-fragmentation diagram shown in figure~\ref{figpu}\,(a). The propagator
for the exchanged gluon has the scaling $1/p_T^2$. The gluon propagators
attached to the $c\bar{c}$ and $b\bar{b}$ pairs are of order $1/m_{J/\psi}^2$
and $1/m_\Upsilon^2$, respectively. The scaling of the product of the two 
triple-gluon vertices must be the square of the typical momentum squared
$p_T^2$. Therefore, the resultant relative scaling of the double-fragmentation
process is $v_c^4v_b^4/(m_{J/\psi}^2m_\Upsilon^2)$, where we have included
the VSR suppression factor $\mathcal{V}=v_c^4 v_b^4$.

In the case of Type A, there are four heavy-quark propagators and one gluon
propagator which have large momentum transfer that account for the scaling
factor $1/p_T^6$. The phase space for the $\mathcal{Q}_1+\mathcal{Q}_2+gg$
final state enhances the scaling factor by an order of $p_T^4$ in comparison
with that for the two-body final state $\mathcal{Q}_1+\mathcal{Q}_2$ of the
double-fragmentation process. As a result, the scaling factor for Type A is
$1/p_T^8$. In a similar manner, we find that the scaling factor for Type B
is the same as that of Type A. In Types A and B, there are two extra hard
jets in the final states, and we have to include the coupling and the 
phase-space suppression factor $\sim g_s^2 /(4\pi)^2 = \alpha_s / (4\pi) $ 
for each hard gluons.  
In addition, the color-octet contribution, which does not have hard jets, 
is distinguished from these color-singlet contributions that can simply 
be removed by imposing an appropriate veto. 
Type C diagrams involve finite box diagrams with two virtual 
gluons whose momentum must be of order the typical momentum transfer $p_T$.
A rough estimate of the scaling can be found by substituting a typical momentum
transfer $p_T$ to the two gluon propagators, four heavy-quark propagators, and
the measure of the loop momentum. Furthermore, the one-loop amplitude has an
additional suppression factor of $g_s^2 /(4\pi)^2 = \alpha_s/(4\pi)$. 
Therefore, the suppression factor for the color-singlet contribution relative
to the color-octet double-fragmentation process is
\begin{equation}
\frac{1}{(4\pi)^2}~\frac{\alpha_s^2}{ v_{c}^{4} v_b^{4}}
\left( \frac{m_{J/\psi}}{p_T} \frac{m_\Upsilon}{p_T} \right)^4,
\label{relsize}%
\end{equation}
where we have included the strong-coupling suppression factor $\alpha^2_s$
to the color-singlet contribution. Therefore one can argue that the
color-singlet contributions to $pp\rightarrow J/\psi + \Upsilon +X $ will
be suppressed enough compared to the color-octet ones.

However, if $p_T$ is small, then the factors $p_T$ in this scaling should
be of order $m_H$ and, therefore,
the suppression factors for the color-singlet contribution relative
to the mixed contributions $c\bar{c}_1({}^3S_1)+b\bar{b}_8({}^3S_1)$ 
and $c\bar{c}_8({}^3S_1)+b\bar{b}_1({}^3S_1)$ are approximately
estimated to be $\sim \alpha_s^2/[(4\pi)^2 v_b^4]$ or
$\alpha_s^2/[(4\pi)^2 v_c^4]$ in the small $p_T$ region.
These factors are much less than order 1 if we assume 
$v_c^2 \sim 0.3$ and $v_b^2 \sim 0.1$ 
and we use the renormalization scale of order $m_b$.
Thus we may conclude that the color-singlet contribution is suppressed
compared to the color-octet contribution coming from the mixed channels
even in the small $p_T$ region.
This rough estimate might fail as $p_T$ approaches to zero.
Then, our prediction 
for the $c\bar{c}_8({}^3S_1)+b\bar{b}_8({}^3S_1)$ contribution
may be contaminated by
order-$\alpha_s^6$ color-singlet contribution at small $p_T$.
The problem
can be resolved by introducing a lower $p_T$ cut $p_T\gtrsim$ 5\,GeV,
where the double fragmentation rises up. The suppression of the
color-singlet contribution to $pp\to J/\psi + \Upsilon+X$ even in moderate
values of $p_T$ has not yet been predicted, because previous analyses with
the fragmentation approximation are valid only for the large-$p_T$ region. 
In addition, the color-singlet contribution is easily distinguishable by 
imposing a veto that the final state must not include hard jets. This is the
main motivation for investigating $pp\to J/\psi + \Upsilon+X$ in this work
as a clean probe to the color-octet mechanism at the LHC. 
This point has been recently reported in ref.~\cite{Ko:2010vh}.
\subsection{Gluon-fragmentation approximation}
In inclusive single-hadron production in $pp$ collisions, a further
factorization happens if $p_T$ is sufficiently large \cite{Collins:1981uk}.
In the case of the ${}^3S_1$ heavy-quarkonium production, the
fragmentation of a gluon into a $Q\bar{Q}_8({}^3S_1)$ pair dominates the
production rate at large $p_T$ \cite{Braaten:1994vv}. If the two
quarkonia produced in $pp\to H_1+H_2+X$ both have large transverse momenta,
then one can guess that such a factorization can be generalized so that
one might be able to write the inclusive double-quarkonium production cross
section in the form:
\begin{equation}
d \sigma_{\rm f}[pp\to H_1+H_2+X] 
=
f_{g/p}\otimes f_{g/p}
\otimes
d \hat{\sigma}[{gg\to gg}]
\otimes
D_{g \to H_1}
\otimes
D_{g \to H_2}
,
\label{xsecgluon}%
\end{equation}
where the subscript in $d \sigma_{\rm f}$ indicates the fragmentation
approximation, $d\hat{\sigma}[gg\to gg]$ is the parton-level cross section
for $gg\to gg$, and $D_{g \to H}(z,\mu_{\rm f})$ is the fragmentation
function for a gluon to fragment into a $Q\bar{Q}_8({}^3S_1)$ pair that
evolves asymptotically into the heavy quarkonium $H=J/\psi$ or $\Upsilon$.
Here, $z$ is the longitudinal momentum fraction of $H$ relative to the
fragmenting gluon and $\mu_{\rm f}$ is the factorization scale for the
fragmentation. An educated guess is that the factorization scale
$\mu_{{\rm f}i}$ for the fragmentation of a gluon into the quarkonium
$H_i$ can be chosen to be of order the transverse mass
$m_{Ti}=(m_i^2+p_{Ti}^2)^{1/2}$, where $m_i$ and $p_{Ti}$ are the
mass and the transverse momentum of $H_i$, respectively. 

At the threshold $\mu_{\rm f}=2m_Q$, $D_{g \to H}(z,\mu_{\rm f}=2m_Q)$
is calculable perturbatively after factoring out the NRQCD matrix element 
$\langle O_8^H({}^3S_1)\rangle$. The fragmentation function of a gluon
that fragments into the ${}^3S_1$ quarkonium via $Q\bar{Q}_8({}^3S_1)$
pair is known up to NLO in $\alpha_s$ at the 
threshold \cite{Braaten:2000pc,Lee:2005jw}. The expression at LO in
$\alpha_s$ is given by~\cite{BBL}
\begin{equation}
D_{g \to H} (z, 2m_Q) =
\frac{\pi \alpha_s}{24 m_Q^3} \delta(1-z)
\langle O_8^H({}^3S_1)\rangle.
\label{frag}%
\end{equation}
In order to take into account multiple emissions of collinear gluons
before creating a $Q\bar{Q}$ pair, it is necessary to evaluate the
fragmentation function at the factorization scale $\mu_{\rm f}\sim p_T\gg m_Q$.
This step can be carried out by making use of the Altarelli-Parisi evolution
equation~\cite{Gribov:1972ri,Altarelli:1977zs,Dokshitzer:1977sg}.

Very recently, Qiao, Sun, and Sun~\cite{Qiao:2009kg} considered the inclusive
process $pp\to 2J/\psi+X$ and Li, Zhang, and Chao~\cite{Li:2009ug} carried out
extensive studies of various double-quarkonium production processes that
include $pp\to 2J/\psi+X$ and $2\Upsilon+X$, where the gluon-gluon initial
states were mainly considered as are in this work. While the color-single
contribution was computed by using the same way that we employ here, the authors
of these papers used the gluon-fragmentation approximation to estimate the
color-octet contributions to $pp\to 2J/\psi+X$~\cite{Li:2009ug,Qiao:2009kg} and 
$2\Upsilon+X$~\cite{Li:2009ug}\footnote{Note that our analysis on the process
$pp\to J/\psi+\Upsilon+X$ is the first NRQCD-based study including both
color-singlet and -octet channels that does not employ the gluon-fragmentation
approximation, while the authors of ref.~\cite{Barger:1979js} also have
commented about the process at low-energy hadron collisions by making use
of the color-evaporation model.}. However, in refs.~\cite{Li:2009ug,Qiao:2009kg}
the fragmentation function $D_{g \to H}(z,\mu_{\rm f})$ was not evaluated at
$\mu_{\rm f}\sim p_T$ but at $\mu_{\rm f}=2m_Q$. In fact, in
order to make the fragmentation approximation valid, the fragmenting gluon has
energies of order $p_T\gg m_Q$ where the strong coupling is significantly
smaller. Therefore, one may guess that the predictions given in
refs.~\cite{Li:2009ug,Qiao:2009kg} must have been overestimated.
On the other hand, once we employ the gluon-fragmentation approximation
to compute the color-octet contributions, the prediction cannot be 
extended to a lower $p_T$ region, where the octet contribution acquires
IR divergences. At small $p_T$, the fragmentation contribution loses
its predictive power. Therefore, the predictions in
refs.~\cite{Li:2009ug,Qiao:2009kg} do not extend to lower values of $p_T$.
Thus, our choice of not employing the gluon-fragmentation approximation
in this work enables us to predict the $p_T$ spectrum to a wider
range whose lower bound extends to $p_T=0$.
\subsection{QED contribution}
As another color-singlet contribution to the double-quarkonium production
at hadron colliders, one may consider the parton-level process 
$q\bar{q}\to Q\bar{Q}_1({ }^3S_1)+Q\bar{Q}_1({ }^3S_1)$.
This subprocess may acquire a large kinematic enhancement factor due
to the double photon fragmentation, $q\bar{q}\to \gamma^*\gamma^*$
followed by $\gamma^*\to Q\bar{Q}_1({ }^3S_1)$ for each virtual
photon. As the double-$J/\psi$ production in $e^+e^-$ annihilation
into two virtual photons, this process has additional logarithmic
enhancement in the forward region~\cite{double2}. 
However, in hadroproduction,
it is very difficult to analyze that kinematic region. In addition, the 
rough estimate of this QED contribution to the double-gluon-fragmentation
contribution is $\sim e_q^4 e_c^2 e_b^2 \alpha^4/(\alpha_s^4 v_c^4 v_b^4)
\sim O(10^{-5})$ with the electromagnetic coupling $\alpha$
and electric charges $e_q$, $e_c$, and $e_b$ of the initial,
charm, and bottom quark, respectively, which is negligible numerically.
This $q\bar{q}$-initiated QED contribution is further suppressed compared
to the gluon-initiated processes. Therefore, we do not consider the QED
contribution in this work. 
\section{Numerical Analysis\label{sec:numerical}}
Based on the formalism described in section~\ref{sec:strategy}, we are ready
to carry out the numerical calculation of the production rates for the
double-quarkonium production processes $pp\to 2J/\psi+X$, $2\Upsilon+X$,
and $J/\psi+\Upsilon+X$ at the LHC with the CM energies $\sqrt{s}=$ 7
and 14\,TeV. In this section, we describe our choice of various input
parameters and the strategies of the numerical evaluation of these cross
sections.
\subsection{Parameters involving NRQCD factorization}
In order to evaluate the production rates in eq.~(\ref{xsec}), we have to
know the values for the long-distance NRQCD matrix elements
$\langle O_n^{H}(^3S_1) \rangle$ for $H=J/\psi$ and $\Upsilon$, where
$n=1$ or 8. The color-singlet matrix element $\langle O_1(^3S_1) \rangle_H$
for the spin-triplet $S$-wave heavy quarkonium decay is usually determined
from the leptonic decay rate of $H$, which is the most precisely measured
value involving $\langle O_n(^3S_1) \rangle_H$%
\footnote{There has been a great progress in precise
determination of the color-singlet matrix element of the $J/\psi$ meson
due to the introduction of a new technique to deal with the relativistic
corrections and its resummation in conjunction with the one-loop QCD
correction to the electromagnetic decay rate \cite{Bodwin:2006dn,%
Bodwin:2006dm,Bodwin:2007fz,Bodwin:2008vp,Chung:2008sm}.
An analogous method has been used to determine the matrix elements for
corresponding bottomonium states \cite{Kang:2007uv}.}.
Under the vacuum-saturation approximation, the decay matrix element is
approximately a third of the production matrix element up to corrections
of order $v_Q^4$ \cite{BBL}:
$\langle O_n^{H}(^3S_1) \rangle=3\langle O_1(^3S_1) \rangle_H+O(v_Q^4)$,
where the factor of 3 on the right side of the equality stands for the
spin-multiplicity factor $(2J+1)$ for the $S$-wave spin-triplet state.
We quote the following values for the color-singlet NRQCD matrix elements
for $J/\psi$ and $\Upsilon$:
\begin{subequations}
\begin{eqnarray}
  \langle O_1^{J/\psi}(^3S_1) \rangle &=& 1.3
\,{\rm GeV}^3~\mbox{\cite{Bodwin:2007fz}},
\\
\langle O_1^{\Upsilon} (^3S_1) \rangle &=& 9.2
\,{\rm GeV}^3~\mbox{\cite{Kang:2007uv}}.
\end{eqnarray}
\end{subequations}
The color-octet matrix element $\langle O_8^{J/\psi} (^3S_1) \rangle$
has been fit to the $p_T$ spectrum of prompt $J/\psi$
production rate at the Tevatron in the large-$p_T$ region~\cite{Braaten:1999qk}.
The matrix element $\langle O_8^{\Upsilon} (^3S_1) \rangle$ has also 
been fit \cite{Braaten:2000cm} to the Tevatron data and used for the
polarization analysis \cite{Braaten:2000gw}. Various determinations
of these matrix elements can be found, for example, in 
ref.~\cite{Kramer:2001hh}. We employ the following values for the
${}^3S_1$ color-octet NRQCD matrix elements for $J/\psi$ and $\Upsilon$:
\begin{subequations}
\begin{eqnarray}
\langle O_8^{J/\psi} (^3S_1) \rangle &=& 3.9\times 10^{-3}
\,{\rm GeV}^3~\mbox{\cite{Braaten:1999qk}},
\\
\langle O_8^{\Upsilon} (^3S_1) \rangle &=& 1.5\times 10^{-1}
\,{\rm GeV}^3~\mbox{\cite{Kramer:2001hh}}.
\end{eqnarray}
\end{subequations}

The short-distance coefficient, which is the parton-level differential
cross section $d\hat{\sigma}$ in eq.~(\ref{xsec}), depends on the
heavy-quark mass $m_Q$ and the strong coupling $\alpha_s$. For $m_Q$,
we take $m_c=1.5$ GeV for the charm quark and $m_b=4.7$ GeV for the
bottom quark. In the NRQCD factorization formula, the short-distance
coefficients are expressed in terms of $m_Q$ instead of the meson mass
$m_H$. This may give rise to relativistic corrections due to the
difference between $m_H$ and $2m_Q$ at higher orders in $v_Q$.
However, because we carry out our calculation at LO in $v_Q$, we
can ignore such corrections and put $m_H=2m_Q$ in our analysis.
\subsection{Other input parameters}
In addition to the NRQCD factorization, the cross section formula
(\ref{xsec}) involves the factorization of the long-distance PDF's,
$f_{a/p}$ and $f_{b/p}$, and the short-distance parton-level cross section 
$d\hat{\sigma}$ with the factorization scale $\mu$. For the scale $\mu$,
we take the transverse mass $\mu= m_T = ( 4 m_Q^2 + p_T^2)^{1/2}$, which
is a conventional choice\footnote{See, for example, 
ref.~\cite{Braaten:1999qk}.}. In general, the transverse momenta for the
two heavy quarkonia in the final state can be different. However, at LO
in $\alpha_s$, there are no additional hard jets and the two final-state 
quarkonium pairs have the same $p_T$ and, therefore, $p_T$ and $m_T$ are
defined unambiguously. As a specific choice of the PDF in eq.~(\ref{xsec}),
we employ the CTEQ6L parametrization~\cite{Pumplin:2002vw}.

In evaluating $\alpha_s$, we set the renormalization scale to be $m_T$
so that $\alpha_s=\alpha_s(\mu=m_T)$. In order to make our numerical
evaluation of $\alpha_s$ consistent with the CTEQ6L parametrization,
we use the NLO formula for the running coupling constant $\alpha_s(\mu)$
by setting $\alpha_s(\mu=M_Z)=0.118$ with 
$\Lambda_4=326$\,MeV~\cite{Pumplin:2002vw}.

The kinematic region that we study in this work covers the rapidity range 
$|y|<2.4$. Our results for the color-octet contributions are finite as well
as the color-singlet contribution at $p_T=0$, since we do not adopt the
gluon-fragmentation approximation. Therefore, we set the lower bound
of $p_T$ to be $0$ that is distinguished from the previous studies in 
refs.~\cite{Li:2009ug,Qiao:2009kg}, where they considered the range
$3\,\textrm{GeV}<p_T<50$\,GeV.
\section{Predictions\label{sec:predictions}}
In this section, we provide our predictions of the differential
cross sections for the double-quarkonium productions
$pp\to 2J/\psi+X$, $2\Upsilon+X$, and $J/\psi+\Upsilon+X$
at CM energies $\sqrt{s}=$ 7 and $14$\,TeV.
Detailed strategies and the values for the input parameters 
for the numerical analysis have been given in sections
\ref{sec:strategy} and \ref{sec:numerical}, respectively.

\begin{figure}
\begin{center}
\begin{tabular}{cc}
\epsfig{file=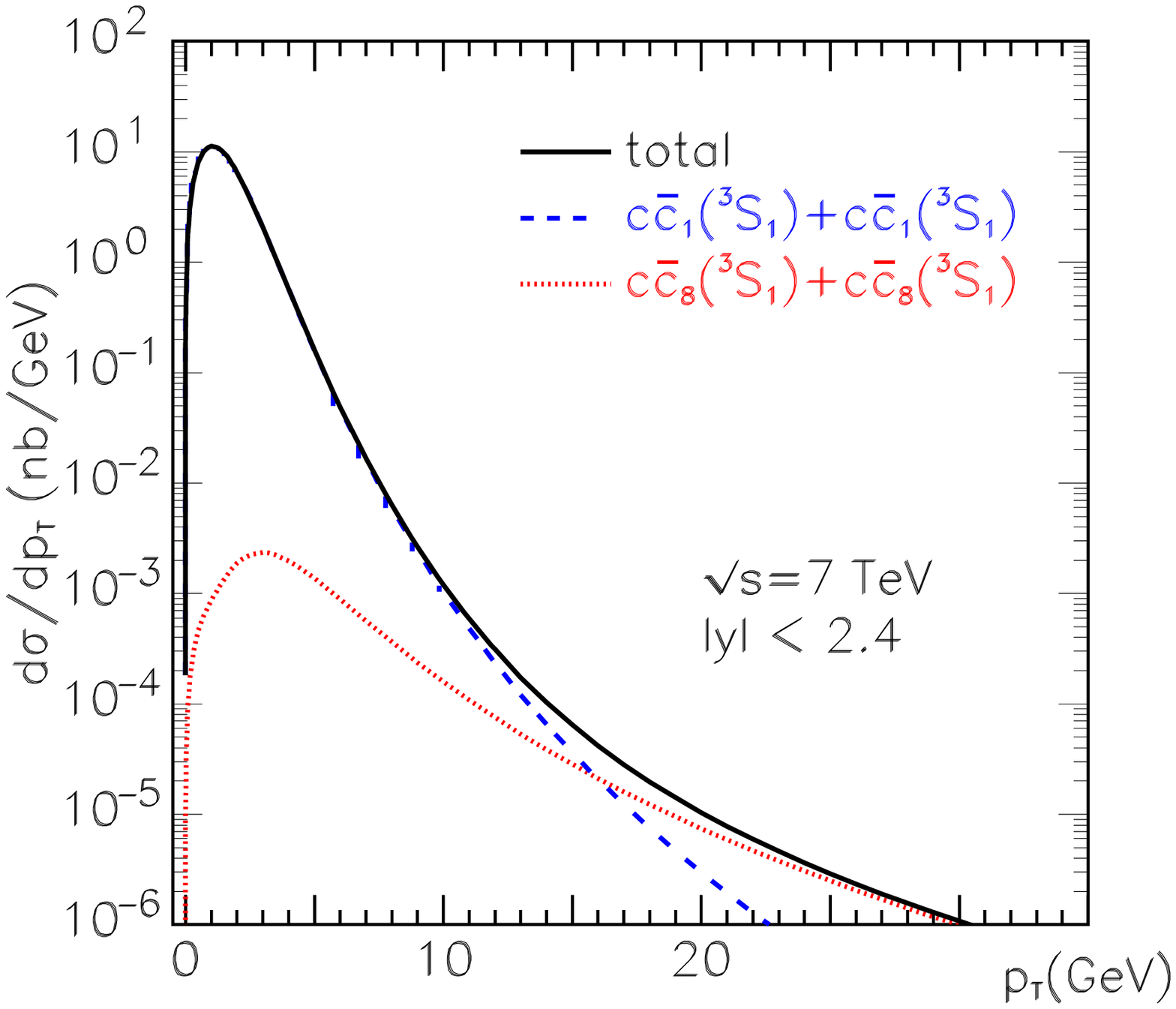,width=0.48\textwidth}
&
\epsfig{file=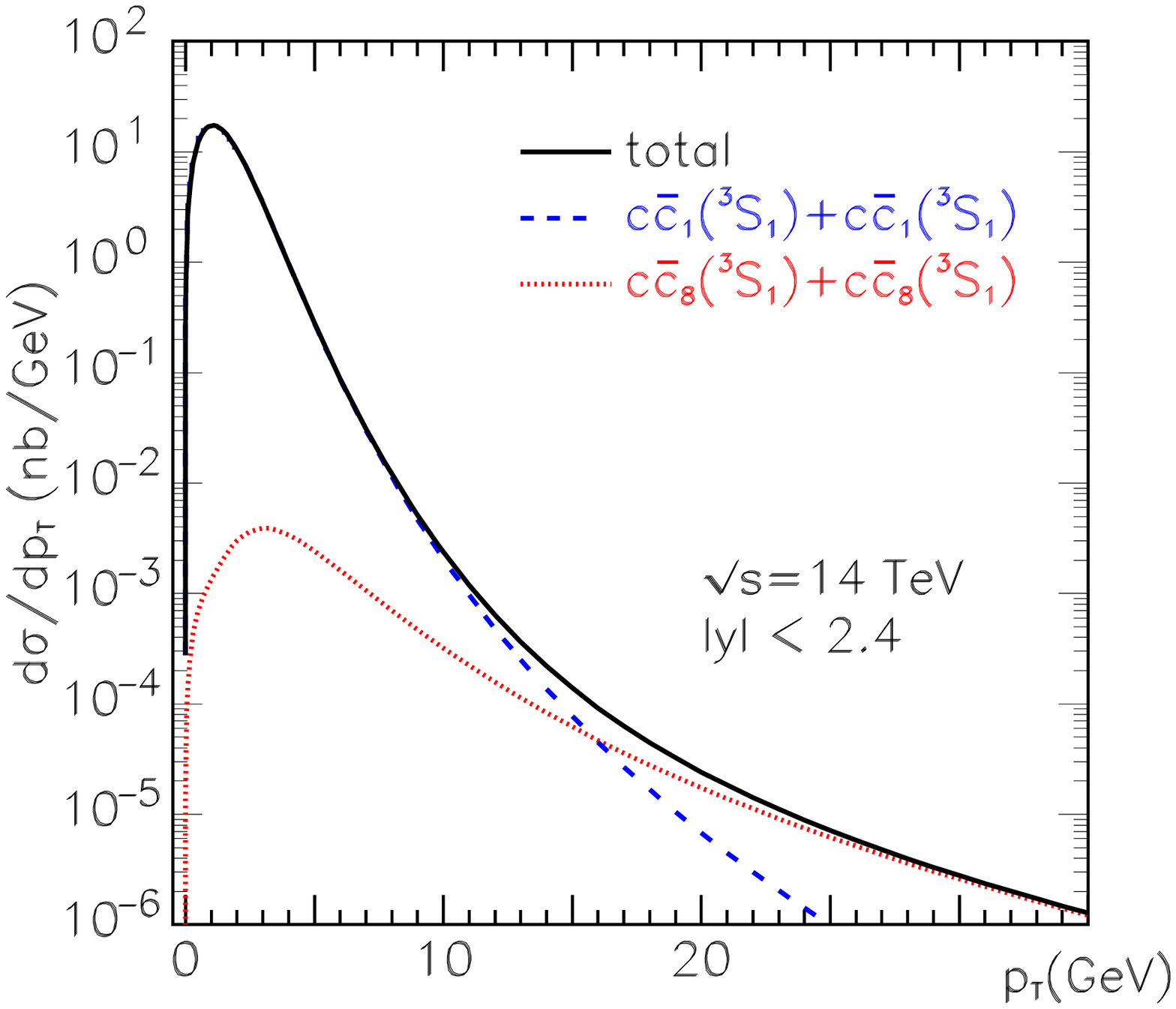,width=0.48\textwidth}
\\[-8.5ex]
\textsf{(a)}&\textsf{(b)}
\end{tabular}
\caption{\label{fig:psi}%
The differential cross sections for $pp\to 2J/\psi+X$
at (a) $\sqrt{s}=7$\,TeV and (b) 14\,TeV 
in units of nb/GeV as functions of $p_T$ integrated over
the rapidity range $|y|<2.4$.
The solid, dashed, and dotted curves represent
the total, color-singlet [$c\bar{c}_1({}^3S_1)+c\bar{c}_1({}^3S_1)$], 
and color-octet [$c\bar{c}_8({}^3S_1)+c\bar{c}_8({}^3S_1)$] 
contributions, respectively.}
\end{center}
\end{figure}

\begin{figure}
\begin{center}
\begin{tabular}{cc}
\epsfig{file=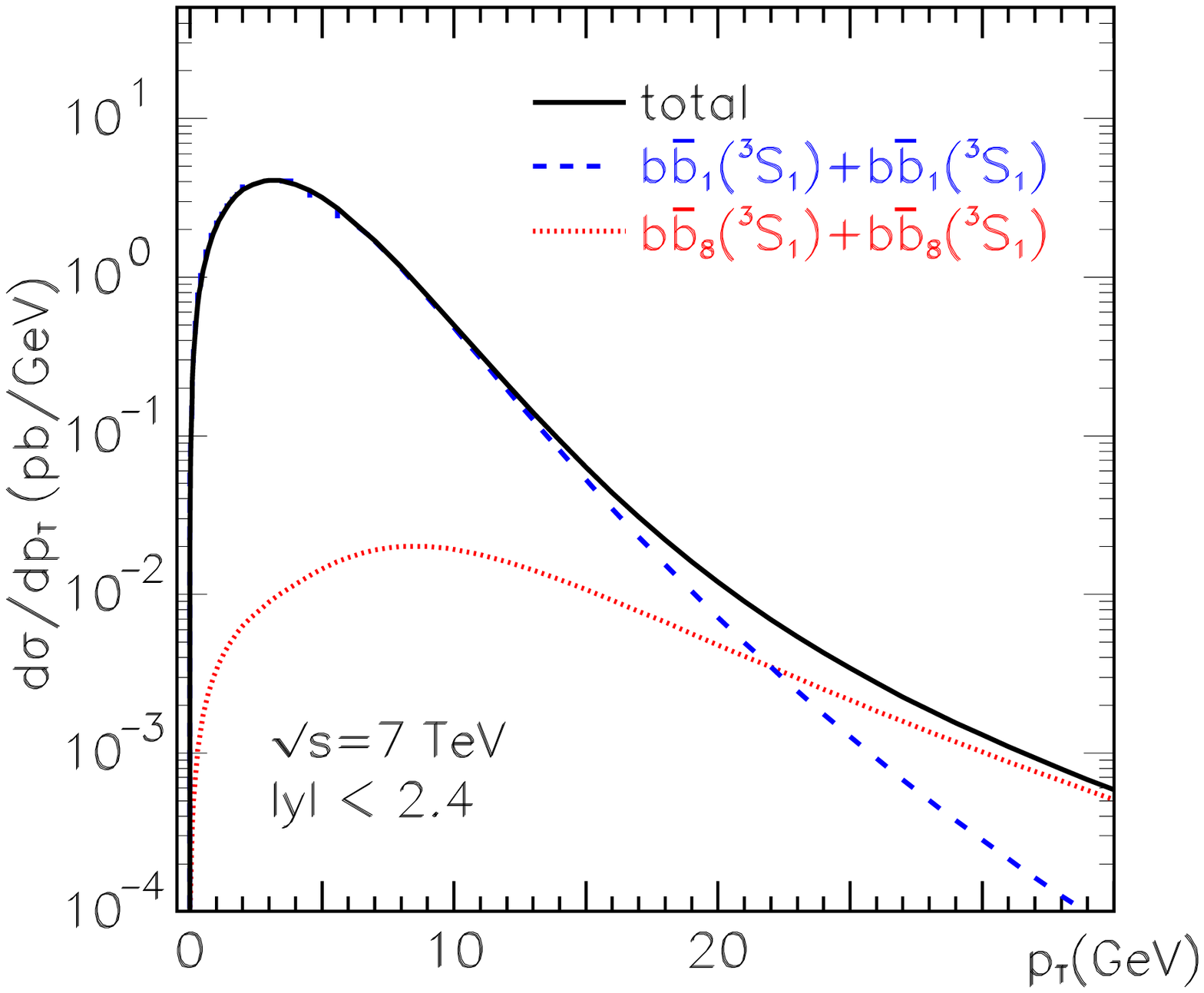,width=0.48\textwidth}
&
\epsfig{file=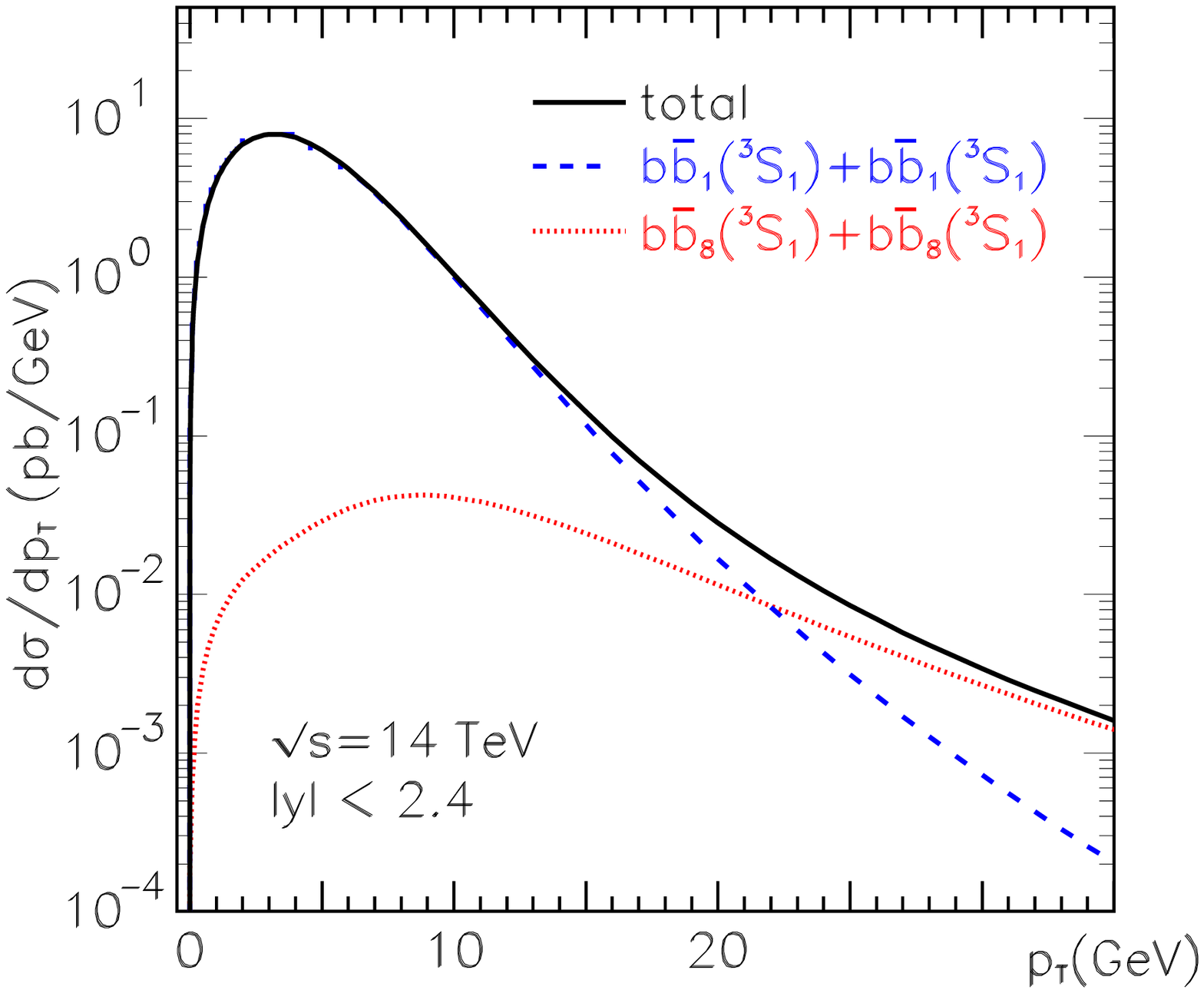,width=0.48\textwidth}
\\[-8.5ex]
\textsf{(a)}&\textsf{(b)}
\end{tabular}
\caption{\label{fig:ups}%
The differential cross sections for $pp\to 2\Upsilon+X$
at (a) $\sqrt{s}=7$\,TeV and (b) 14\,TeV 
in units of pb/GeV as functions of $p_T$ integrated over
the rapidity range $|y|<2.4$.
The solid, dashed, and dotted curves represent
the total, color-singlet [$b\bar{b}_1({}^3S_1)+b\bar{b}_1({}^3S_1)$], 
and color-octet [$b\bar{b}_8({}^3S_1)+b\bar{b}_8({}^3S_1)$] 
contributions, respectively.}
\end{center}
\end{figure}

\subsection{$d\sigma/dp_T$ for $pp\to 2J/\psi+X$ 
and $2\Upsilon+X$\label{sec:spect2H}}
Our predictions for the $p_T$ dependence of the differential cross sections 
$d\sigma/dp_T[pp\to 2J/\psi+X]$ at $\sqrt{s}=$ 7 and 14\,TeV  integrated
over the rapidity range $|y|<2.4$ are shown as solid curves in 
figures~\ref{fig:psi}\,(a) and (b), respectively, and those for
$d\sigma/dp_T[pp\to 2\Upsilon+X]$ are shown in figure~\ref{fig:ups}. As is
stated in section~\ref{sec:ii-identical}, we have considered only
order-$\alpha_s^4$ gluon-initiated parton processes: the color-singlet
contribution through $gg\to Q\bar{Q}_1({}^3S_1)+Q\bar{Q}_1({}^3S_1)$
and the color-octet contribution 
$gg\to Q\bar{Q}_8({}^3S_1)+Q\bar{Q}_8({}^3S_1)$ for these
identical-quarkonium pair productions. 

As far as the shape is concerned, the $p_T$ spectra at the two CM energies
$\sqrt{s}=$ 7 and 14\,TeV are essentially the same. The ratio of the
value at $\sqrt{s}=14$\,TeV to that at $\sqrt{s}=7$\,TeV is about a factor
of $1.5$ ($2$) for $pp\to 2J/\psi+X$ ($2\Upsilon+X$) near the peak, where
the color-singlet contribution dominates. As $p_T$ increases, the ratio
increases and the gluon-fragmentation contribution dominates. The $p_T$
spectra in figures~\ref{fig:psi} and \ref{fig:ups} show that the color-singlet
channel dominates at small $p_T$ while the color-octet channel dominates at
large $p_T$. The crossovers are placed at $p_T\approx 16$\,GeV for
$pp\to 2 J/\psi+X$ and at $p_T\approx 24$\,GeV for $pp\to 2 \Upsilon+X$
at $\sqrt{s}=$ 7 and 14\,TeV both. Another noticeable feature
of the spectra is that the values at $p_T=0$ vanish:
\begin{equation}
\lim_{p_T\to 0}\frac{d\sigma}{dp_T}[pp\to 2H+X]=0.
\end{equation}
The reason is that, at order $\alpha_s^4v_Q^0$, all of the diagrams we
consider in this work do not contain any real gluons in the final state and
both color-singlet and -octet contributions are free of IR
divergences\footnote{However, if there is at least one real gluon in the
final state, then in the limit that the final gluons have vanishing momenta,
IR divergences may appear in the  color-octet contributions even at LO
$v_Q^0$. In the case of the color-singlet contributions such an IR
divergence may grow at higher orders in $v_Q^n$ even at LO in $\alpha_s$.
For more discussions, we refer the readers to 
refs.~\cite{Bodwin:2002hg,Bodwin:2003wh,Bodwin:2009cb,Bodwin:2010fi}.}.
In the low-$p_T$ region, the differential cross sections increase as $p_T$
increases until they reach the maximum values
$d\sigma/dp_T[pp\to 2J/\psi+X]=11.2$ (17.3)\,nb/GeV at
$p_T=1.1$\,GeV and
$d\sigma/dp_T[pp\to 2\Upsilon+X]=4.1$ (8.0)\,pb/GeV at
$p_T=3$ \,GeV for $\sqrt{s}=7$ (14)\,TeV.
\begin{table}[th]
\begin{center}
\begin{tabular}{c|ccc}
\hline
\hline
~~$\sqrt{s}$ $\backslash$ $\sigma$ (nb)~~
& ~~$c\bar{c}_1({}^3S_1)+c\bar{c}_1({}^3S_1)$~~
& ~~$c\bar{c}_8({}^3S_1)+c\bar{c}_8({}^3S_1)$~~
& ~~total~~
\\
\hline
\phantom{1}7 TeV & 22.3 &  0.011 & 22.3 
\\
14 TeV & 34.8 & 0.019 & 34.8 
\\
\hline
\hline
\end{tabular}
\end{center}
\caption{\label{tablepsi}%
The total cross sections $\sigma[pp\to 2J/\psi+X]$ at
$\sqrt{s}=7$ and $14$\,TeV integrated over the ranges
$|y|<2.4$ and $p_T<50$\,GeV in units of nb.
The three columns represent the color-singlet, color-octet,
and total contributions.}
\end{table}
\begin{table}[th]
\begin{center}
\begin{tabular}{c|ccc}
\hline
\hline
~~$\sqrt{s}$ $\backslash$ $\sigma$ (pb)~~
& ~~$b\bar{b}_1({}^3S_1)+b\bar{b}_1({}^3S_1)$~~
& ~~$b\bar{b}_8({}^3S_1)+b\bar{b}_8({}^3S_1)$~~
& ~~total~~
\\
\hline
\phantom{1}7 TeV & 24.1 &  0.27 & 24.4
\\
14 TeV & 47.9 & 0.60 & 48.5
\\
\hline
\hline
\end{tabular}
\end{center}
\caption{\label{tableups}%
The total cross sections $\sigma[pp\to 2\Upsilon+X]$ at
$\sqrt{s}=7$ and $14$\,TeV integrated over the ranges
$|y|<2.4$ and $p_T<50$\,GeV in units of pb.
The three columns represent the color-singlet, color-octet,
and total contributions.
}
\end{table}

Next we compute the total cross sections $\sigma[pp\to 2J/\psi+X]$ and
$\sigma[pp\to 2\Upsilon+X]$ by integrating eq.~(\ref{xsec}) over the ranges
$|y|<2.4$ and  $p_T<50$\,GeV. As shown in tables~\ref{tablepsi} and 
\ref{tableups}, they are $\sigma[pp\to 2J/\psi+X]=22$ (35)\, nb 
and $\sigma[pp\to 2\Upsilon+X]=24$ (49)\, pb at $\sqrt{s}=7$ (14)\,TeV.
Because the differential cross section is dominated by the color-singlet
channel near the peak, the total cross section is dominated by
the color-singlet contribution.
\begin{figure}[th]
\begin{center}
\begin{tabular}{cc}
\epsfig{file=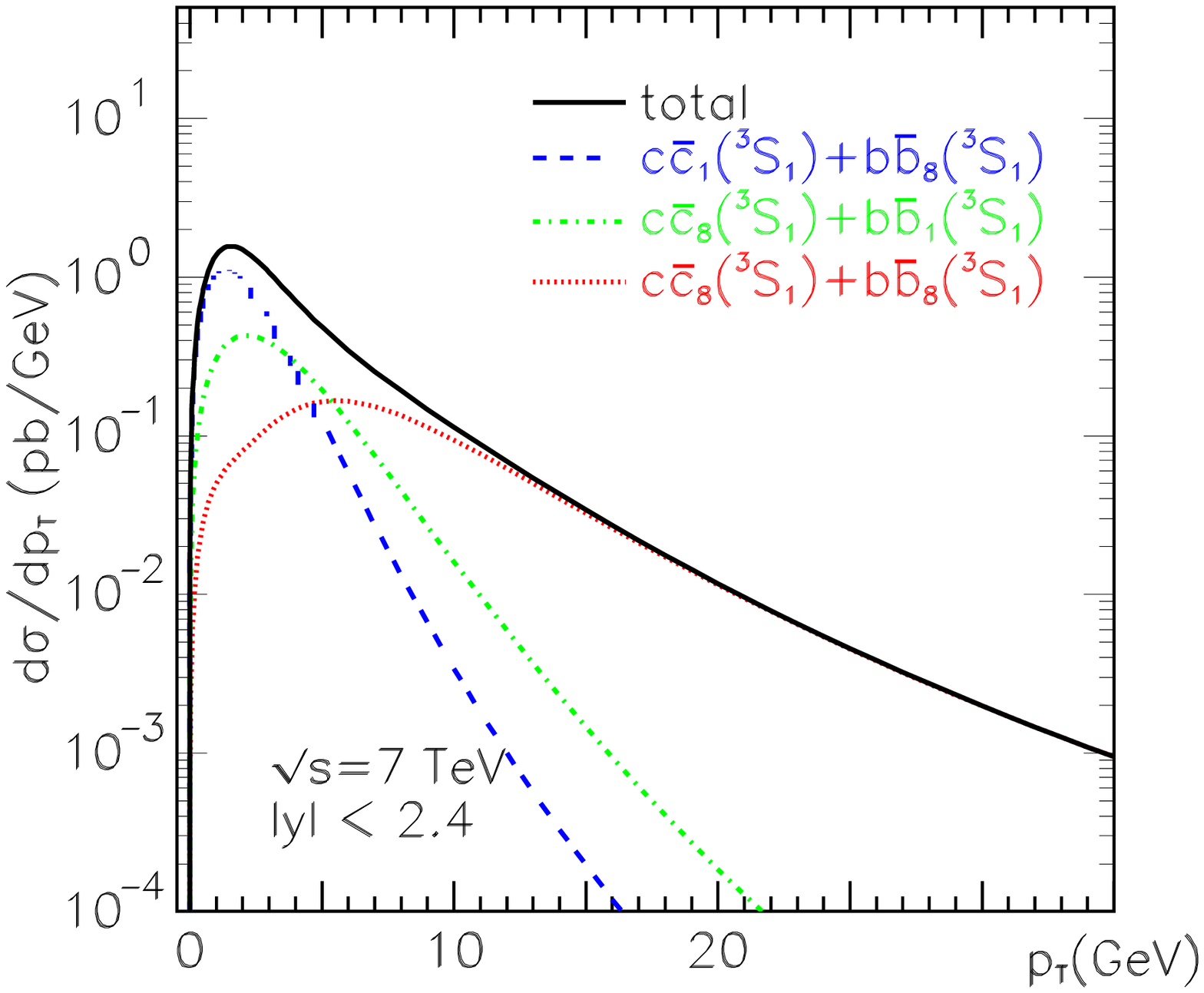,width=0.48\textwidth}
&
\epsfig{file=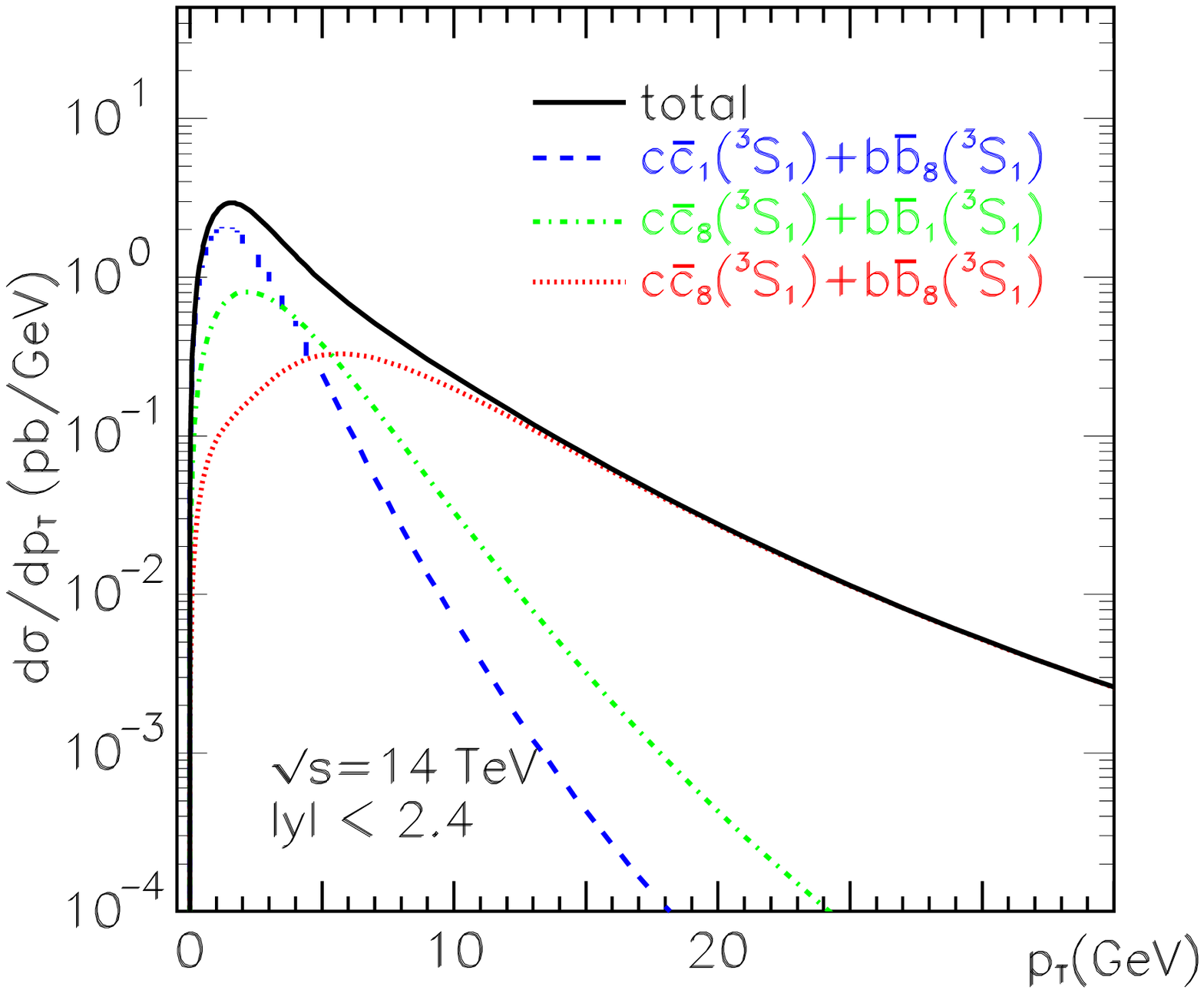,width=0.48\textwidth}
\\[-8.5ex]
\textsf{(a)}&\textsf{(b)}
\end{tabular}
\caption{\label{fig:psi-ups}%
The differential cross sections for $pp\to J/\psi+\Upsilon+X$
at (a) $\sqrt{s}=7$\,TeV and (b) 14\,TeV 
in units of pb/GeV as functions of $p_T$ integrated over
the rapidity range $|y|<2.4$.
The solid, dashed, dashed-dotted, and dotted curves represent
the total, 
$c\bar{c}_1({}^3S_1)+b\bar{b}_8({}^3S_1)$, 
$c\bar{c}_8({}^3S_1)+b\bar{b}_1({}^3S_1)$, and
$c\bar{c}_8({}^3S_1)+b\bar{b}_8({}^3S_1)$
contributions, respectively.}
\end{center}
\end{figure}
\subsection{$d\sigma/dp_T$ for $pp\to J/\psi+\Upsilon+X$%
\label{sec:spectH12}}
The predictions for $d\sigma/dp_T[pp\to J/\psi+\Upsilon+X]$ at
$\sqrt{s}=7$\,TeV and 14\,TeV integrated over the rapidity range $|y|<2.4$
are shown as solid curves in figures~\ref{fig:psi-ups}\,(a) and (b),
respectively. In this process, we consider the following order-$\alpha_s^4$
contributions: the color-octet channel
$gg\to c\bar{c}_8({}^3S_1)+b\bar{b}_8({}^3S_1)$
and the two mixed channels
$gg\to c\bar{c}_1({}^3S_1)+b\bar{b}_8({}^3S_1)$ and
$gg\to c\bar{c}_8({}^3S_1)+b\bar{b}_1({}^3S_1)$.
The reason for neglecting the color-singlet contribution and the remaining
color-octet channels is given in section~\ref{sec:ii-different}.

The $p_T$ spectrum for $pp\to J/\psi+\Upsilon+X$ is similar to those of
$pp\to 2 J/\psi+X$ and $2 \Upsilon+X$. The differential cross section
vanishes at $p_T=0$ and rapidly increases until it reaches the maximum
value $d\sigma/dp_T[pp\to J/\psi+\Upsilon+X]=1.6$ (2.9)\,pb/GeV at 
$p_T=1.6$ (1.5)\,GeV for $\sqrt{s}=7$ (14)\,TeV. Then it monotonically decays
as $p_T$ increases. Near $p_T=0$, the $c\bar{c}_1({}^3S_1)+b\bar{b}_8({}^3S_1)$
channel dominates and that is the largest contribution for
$p_T \lesssim$ 4 \,GeV. At large $p_T$, the color-octet channel
$c\bar{c}_8({}^3S_1)+b\bar{b}_8({}^3S_1)$ dominates
and the channel is the largest contribution for $p_T \gtrsim$ 6 \,GeV.
In the remaining region, about $4 \,\textrm{GeV} \lesssim p_T \lesssim $ 
6 \,GeV, the color-octet channel $c\bar{c}_8({}^3S_1)+b\bar{b}_8({}^3S_1)$
and the two mixed channels
$c\bar{c}_1({}^3S_1)+b\bar{b}_8({}^3S_1)$ and
$c\bar{c}_8({}^3S_1)+b\bar{b}_1({}^3S_1)$ compete together.
\begin{table}[th]
\begin{center}
\begin{tabular}{c|cccc}
\hline
\hline
$\sqrt{s}$ $\backslash$ $\sigma$\,(pb)
& $c\bar{c}_1({}^3S_1)+b\bar{b}_8({}^3S_1)$
& $c\bar{c}_8({}^3S_1)+b\bar{b}_1({}^3S_1)$
& $c\bar{c}_8({}^3S_1)+b\bar{b}_8({}^3S_1)$
& total
\\
\hline
7 TeV & 3.18 &  1.95 & 1.63 & \phantom{1}6.76
\\
14 TeV & 6.00 & 3.72 & 3.36 & 13.08
\\
\hline
\hline
\end{tabular}
\end{center}
\caption{\label{tablepsiups}%
The total cross sections $\sigma[pp\to J/\psi+\Upsilon+X]$ at
$\sqrt{s}=7$ and $14$\,TeV integrated over the ranges
$|y|<2.4$ and $p_T<50$\,GeV in units of pb.
The four columns represent the
$c\bar{c}_1({}^3S_1)+b\bar{b}_8({}^3S_1)$,
$c\bar{c}_8({}^3S_1)+b\bar{b}_1({}^3S_1)$,
$c\bar{c}_8({}^3S_1)+b\bar{b}_8({}^3S_1)$, and
total contributions.
}
\end{table}

We compute the total cross section $\sigma[pp\to J/\psi+\Upsilon+X]$
by integrating eq.~(\ref{xsec}) over the ranges $|y|<2.4$ and $p_T<50$\,GeV.
As shown in table~\ref{tablepsiups}, the total cross section at
$\sqrt{s}=7\,(14)$\,TeV is $\sigma[pp\to J/\psi+\Upsilon+X]=6.8\,(13)$\,pb.
The total cross section for $pp\to J/\psi+\Upsilon+X$ without any $p_T$  cut
is smaller than those for $2J/\psi$ and $2\Upsilon$ final states by
factors of $\sim 3000$ and $\sim 4$, respectively. We notice that most
of the rates for the $2J/\psi$ and $2\Upsilon$ cases are concentrated in
the low $p_T$ regions, where the color-singlet contribution dominates.
In order to probe the color-octet contribution more accurately, we had better
impose a lower $p_T$ cut to remove most of the color-singlet contribution.
The cut for the $2J/\psi$ or $2\Upsilon$ case should be around the place
where the $Q\bar{Q}_8({}^3S_1)+Q\bar{Q}_8({}^3S_1)$ contribution overtakes
the $Q\bar{Q}_1({}^3S_1)+Q\bar{Q}_1({}^3S_1)$ contribution. According to
figures~\ref{fig:psi} and \ref{fig:ups}, the cuts $p_T\gtrsim 16$ and 
24\,GeV are appropriate for $2J/\psi$ and $2\Upsilon$ final states,
respectively. In those cases, 
$\sigma[pp\to 2J/\psi+X]|_{p_T\gtrsim 16\,{\rm GeV}}=0.09\,(0.2)$\,pb,
$\sigma[pp\to 2\Upsilon+X]|_{p_T\gtrsim 24\,{\rm GeV}}=0.02\,(0.05)$\,pb,
respectively, at $\sqrt{s}=$ 7\,(14)\,TeV.
As is described in section~\ref{sec:ii-different},
the whole cross section $\sigma[pp\to J/\psi+\Upsilon+X]=6.8\,(13)$\,pb
depends on the color-octet matrix elements and there is no need to impose
an additional $p_T$ cut to remove the color-singlet contribution.
Therefore, we conclude that the $pp\to J/\psi+\Upsilon+X$ channel is
the most sensitive to the color-octet matrix elements among the three
double-quarkonium final states.
\subsection{Comparison with previous calculations}
We compare our predictions in sections~\ref{sec:spect2H} and \ref{sec:spectH12}
with previous calculations \cite{Li:2009ug,Qiao:2009kg} based on the
gluon-fragmentation approximation. We find that our prediction is
distinguished from the previous calculations in the shape of the $p_T$ spectrum
and these previous calculations severely overestimate the rates all over the 
range of $p_T.$

First of all, while our $p_T$ spectra 
in figures~\ref{fig:psi}--\ref{fig:psi-ups}
have maximum values around $p_T\approx 2$, $4$--$5$, and $2$--$3$\, GeV
for $pp\to 2J/\psi+X$, $2\Upsilon+X$, and $J/\psi+\Upsilon+X$, respectively, 
the previous predictions in refs.~\cite{Li:2009ug,Qiao:2009kg} do not have
these peaks. The values of $p_T$ at which the peaks appear are of order
$m_{J/\psi}$ for $pp\to 2J/\psi+X$ and $pp\to J/\psi+\Upsilon+X$ and of order
$m_{\Upsilon}$ for $pp\to 2\Upsilon+X$. This is the region where the fusion
contributions  are important. Neglecting fusion diagrams in the fragmentation
approximation cannot reproduce the correct shape of $p_T$ spectrum
at transverse momenta less than a few $m_H$.

In addition, those previous predictions based on the 
gluon-fragmentation approximation overestimate the rates all over
the $p_T$ range. At small $p_T$, the fragmentation approximation must
break down and, therefore, the approximation severely overestimate the
rate in this region. The approximation actually diverges as $p_T\to 0$
while our prediction vanishes in that limit.
In fact, the fragmentation approximation must be a good approximation
if $p_T$ is sufficiently large.
At large $p_T$, the predictions in refs.~\cite{Li:2009ug,Qiao:2009kg},
again, severely overestimate the rate. The reason is that in these
references the authors fixed the factorization scale for the gluon
fragmentation to be $\mu_{\rm f}=2m_Q$, which leads to large cross
sections in comparison with our predictions given in
figures~\ref{fig:psi} and \ref{fig:ups} in which we use conventional
choice $\mu_{\rm f}=m_T$.

We can compare our results for the double-quarkonium production of
the same flavor, $p p \to 2J/\psi+X$ and 
$p p \to 2\Upsilon+X$ with those for the single-quarkonium
production associated with a heavy-quark pair with the same flavor,
$p p \to J/\psi + c\bar{c}+X$ and
$p p \to \Upsilon+ b\bar{b}+X$~\cite{Artoisenet:2007xi}.
In both cases, the color-octet contributions are of the same
order in $\alpha_s$ as those for the color-singlet channels.
Therefore, if there are no kinematic enhancement factors such as
fragmentation,
then the color-octet contributions must be suppressed.
The results shown in figures~\ref{fig:psi} and \ref{fig:ups} are consistent
with this expectation that the color-singlet contribution dominates over
the color-octet contribution unless $p_T$ is extremely large.
Comparing our results for the production rate
with those in ref.~\cite{Artoisenet:2007xi} which contains only 
the color-singlet contribution, we find that the production rate
for the single-quarkonium production associated with a heavy-quark pair
with the same flavor at the LHC is by about a factor of $10^3$ larger than
that for the double-quarkonium production at the LHC, where we have taken into
account the decay of each heavy-quarkonium into a muon pair.
Such a suppression should be due to the restrictions in the phase space
and the bound-state formation. However, the double-quarkonium production
has clear signals, while the single-quarkonium production
associated with an open heavy-quark pair may suffer from large backgrounds.

\section{Discussions\label{sec:discussion}}
Based on the NRQCD factorization formalism, we have obtained the $p_T$
spectra and the total cross sections for $pp\to 2J/\psi+X$, $2\Upsilon+X$,
and $J/\psi+\Upsilon+X$ at CM energies $\sqrt{s}=7$\,TeV and 14\,TeV.
The total production rates integrated over the rapidity and 
transverse-momentum ranges $|y|<2.4$ and $p_T<50$\,GeV are predicted to be
$\sigma[pp\to 2J/\psi+X]=$ 22 (35)\,nb,
$\sigma[pp\to 2\Upsilon+X]=$ 24 (49)\,pb, and
$\sigma[pp\to J/\psi+\Upsilon+X]=$ 7 (13)\,pb at $\sqrt{s}=7$ (14)\,TeV.

For the double-quarkonium production of the same flavor, $pp\to 2J/\psi+X$
and $2\Upsilon+X$, production rates are dominated by the color-singlet
contribution at low $p_T$. In order to probe the color-octet contribution,
which dominates at large $p_T$, more accurately we have imposed lower $p_T$
cuts where the color-octet contribution overtakes the color-singlet one.
The results are
$\sigma[pp\to 2J/\psi+X]|_{p_T\gtrsim 16\,{\rm GeV}}=0.09\,(0.2)$\,pb and
$\sigma[pp\to 2\Upsilon+X]|_{p_T\gtrsim 24\,{\rm GeV}}=0.02\,(0.05)$\,pb.
Assuming the integrated luminosity $\sim 100  \,\textrm{fb}^{-1}$ at
$\sqrt{s}=$14\,TeV and considering the branching fractions
$B[J/\psi\to \mu^+\mu^-]=5.93\,$\% and 
$B[\Upsilon \to \mu^+\mu^-]=2.48\,$\%~\cite{Amsler:2008zzb}, we expect that
approximately 70 (3) double-$J/\psi$ (-$\Upsilon$) events can be observed
by tagging muon pairs under the cuts $p_T\ge 16$ (24)\,GeV
for $pp\to 2J/\psi+X$ $(2\Upsilon+X)$.
The results indicate that it is very difficult to study the color-octet
mechanism by making use of the processes $pp\to 2J/\psi+X$ and
$pp\to 2\Upsilon+X$ at the LHC.

In the case of $pp\to J/\psi+\Upsilon+X$, the color-singlet channel is
absent at LO in $\alpha_s$, and is highly suppressed relative to the
color-octet contributions [ see eq.~(2) ]. The two channels
$c\bar{c}_1({}^3S_1)+b\bar{b}_8({}^3S_1)$ and
$c\bar{c}_8({}^3S_1)+b\bar{b}_8({}^3S_1)$ dominate the production rate at
small and large values of $p_T$, respectively. The contribution
$c\bar{c}_8({}^3S_1)+b\bar{b}_1({}^3S_1)$
is comparable to those two listed above only around the region 
$4\,\textrm{GeV}\lesssim p_T\lesssim 6\,\textrm{GeV}$.
Assuming the integrated luminosity 
$\sim 100  \,\textrm{fb}^{-1}$ at $\sqrt{s}=$14\,TeV and
considering the branching fractions $B[J/\psi\to \mu^+\mu^-]=5.93\,$\% 
and $B[\Upsilon \to \mu^+\mu^-]=2.48\,$\%~\cite{Amsler:2008zzb}, 
we expect that approximately 1900, 520, and 160 events can be observed 
by tagging muon pairs under the cuts $p_T\ge$ 0, 5, and 10\,GeV, 
respectively, at the LHC. 
Improving the acceptances for $J/\psi$ and $\Upsilon$ 
by extensive Monte Carlo studies of final-muon pairs,
one may expect that the observation of the events 
and the determination of the matrix
elements can be quite promising in the near future.
The number of events may be increased by a factor of 4 if one includes
the $e^+ e^-$ decay modes of $J/\psi$ and $\Upsilon$.
It may also be improved by including the subprocess $q\bar{q}\to H_1 + H_2$
via two-gluon exchanges that is neglected in this work.

It is well known that, in inclusive single-quarkonium production
in hadron collisions, there are large corrections at NLO in $\alpha_s$.
For example, the NLO corrections to the color-singlet contribution to 
the inclusive $J/\psi$ production enhance the rate
by an order of magnitude~\cite{Campbell:2007ws,Gong:2008sn,Artoisenet:2008fc}
especially at large $p_T$.
Therefore, one can worry that NLO corrections in $\alpha_s$
may spoil the LO prediction for the $p_T$ spectra presented in this work.
In inclusive single-quarkonium production, NLO subprocesses include
$g g \to J/\psi + g g$, $g g \to J/\psi + q\bar{q}$,
$g q(\bar{q}) \to J/\psi + q g (\bar{q})$,
and $g g \to J/\psi + c\bar{c}$. Although there is a large enhancement from
$t$-channel gluon-exchange diagrams, it is also true that each of new diagrams
has significant contributions, piling up the corrections 
to modify the rate by an order of magnitude.
In the case of double-quarkonium production of the same flavor,
new NLO subprocesses include 
$t$-channel gluon-exchange diagrams such as
$g g \to 2H + g$ and $g q(\bar{q}) \to 2H + q(\bar{q})$, where
$H=J/\psi$ or $\Upsilon$.  But the number of new channels are quite limited 
compared to the single-quarkonium production case. Therefore, we expect that 
the NLO corrections to the color-singlet contribution 
to the double-quarkonium production of the same flavor
may enhance the rate significantly, but not as dramatically as
the single-quarkonium production, while observation of the color-octet
mechanism might be significantly affected. This argument should be tested
by a forthcoming quantitative analysis of the NLO corrections 
to the color-singlet and color-octet contributions.

In the case of $J/\psi+\Upsilon$
production, the suppression factor of the color-singlet channel
to the color-octet one is $\alpha_s^2/p_T^8$ which is significantly
smaller than the corresponding factor $1/p_T^4$ for the single-quarkonium
production, in which the color-singlet and color-octet
contributions are of the same order ($\alpha_s^3$) at LO.
We anticipate that our arguments can be tested quantitatively by measurements
and by explicit calculations of complete order-$\alpha_s^6$ contributions
to the color-singlet channel in near future.

\acknowledgments
The authors would like to thank Kuang-Ta Chao and Rong Li
for their useful comments. They also thank Suyong Choi for providing
valuable information regarding four-muon analysis at the CMS experiment.
They also thank Korea Institute for Advanced Study for providing
computing resources (Abacus System) for this work.
A part of the work (P.K.) was done at Aspen Center for Physics during the
summer workshop in 2010.
The work of P.K. was supported in part by the National Research
Foundation (NRF) through Korea Neutrino Research Center (KNRC)
at Seoul National University.
This work was supported
by Basic Science Research Program through the NRF of Korea funded by the
MEST under Contracts 2009-0072689 (C.Y.) and 2010-0000144 (J.L.).

\appendix
\section{Parton-level cross section
for $g g \to Q \bar{Q}_8({}^3S_1) + Q \bar{Q}_8 ({}^3S_1)$
\label{app:same}}
In this appendix, we present the parton-level differential cross
sections for $pp\to 2 J/\psi+X$ and $pp\to 2\Upsilon+X$. 
We first define the Mandelstam variables for the parton-level process
$g(k_1)g(k_2) \to Q{Q}_{n_1}(p_1) +Q'\bar{Q}^\prime_{n_2}(p_2)$,
where the momentum for each state is given in the parentheses following
the state and $Q{Q}_{n_1}$ and $Q'\bar{Q}^\prime_{n_2}$ evolve into the
quarkonium states $H_1$ and $H_2$, respectively. Each state is on its mass
shell so that $k_1^2=k_2^2=0$, $p_1^2=(2m_Q)^2$, and $p_2^2=(2m_{Q'})^2$.
At LO in $v_Q$ and $v_{Q'}$, the meson masses are set to be $m_{H_1}=2m_Q$
and $m_{H_2}=2m_{Q'}$. If $H_1=H_2$, then $m_{H_1}=m_{H_2}=2m_Q$.
The Mandelstam variables $\hat{s}$, $\hat{t}$, and $\hat{u}$ are defined by
\begin{subequations}
\label{eq:Mandelstam}
\begin{eqnarray}
\hat{s}&=&(k_1+k_2)^2=(p_1+p_2)^2,
\\
\hat{t}&=&(k_1-p_1)^2=(k_2-p_2)^2,
\\
\hat{u}&=&(k_1-p_2)^2=(k_2-p_1)^2.
\end{eqnarray}
\end{subequations}

In $pp\to 2 H+X$ for $H=J/\psi$ or $\Upsilon$ we consider the two channels
$gg \to Q{Q}_1({}^3S_1) + Q{Q}_1({}^3S_1)$ and
$gg \to Q{Q}_8({}^3S_1) + Q{Q}_8({}^3S_1)$, where $Q=c$ ($b$) for 
$H=J/\psi$ ($\Upsilon$). The parton-level cross sections for
$gg \to Q{Q}_1({}^3S_1) + Q{Q}_1({}^3S_1)$ can be found in
refs.~\cite{Li:2009ug,Qiao:2009kg}. That for the subprocess
$gg \to Q{Q}_8({}^3S_1) + Q{Q}_8({}^3S_1)$ is given by
\begin{equation}
\label{eq:sub-2H}
\frac{d \hat{\sigma}}{d \hat{t}} 
[g g \to Q \bar{Q}_8({}^3S_1) + Q \bar{Q}_8 ({}^3S_1)] =
\frac{\pi^3 \alpha_s^4 }
{972 m_{H}^6 \hat{s}^8 (\hat{t}-m_{H}^2)^4 (\hat{u}-m_{H}^2)^4}
\sum_{j=0}^{14} a_{j} m_{H}^{2 j},
\end{equation}
where $m_H=2m_Q$ and the coefficients $a_j$'s are  given by
\begin{eqnarray*}
a_0 &=& 243 \hat{s}^4 \hat{t}^2 (\hat{s}+\hat{t})^2 
        (\hat{s}^2+\hat{s} \hat{t}+\hat{t}^2)^3,
\\
a_{1}&=&-162 \hat{s}^3 \hat{t}^2 (\hat{s}+\hat{t}) 
        (\hat{s}^2+\hat{s} \hat{t}+\hat{t}^2) 
	(9 \hat{s}^5 + 19 \hat{s}^4 \hat{t} + 20 \hat{s}^3 \hat{t}^2 
	 + 7 \hat{s}^2 \hat{t}^3 - 3 \hat{s} \hat{t}^4 - 6 \hat{t}^5),
\\
a_{2}&=& \hat{t} (243 \hat{s}^{11} + 3951 \hat{s}^{10} \hat{t} 
       + 6714 \hat{s}^9 \hat{t}^2 + 14420 \hat{s}^8 \hat{t}^3 
       + 179582 \hat{s}^7 \hat{t}^4 + 919446 \hat{s}^6 \hat{t}^5
\nonumber \\
&&       + 2488136 \hat{s}^5 \hat{t}^6 + 4132862 \hat{s}^4 \hat{t}^7 
         + 4395900 \hat{s}^3 \hat{t}^8 + 2933988 \hat{s}^2 \hat{t}^9 
	 + 1119744 \hat{s} \hat{t}^{10} 
\nonumber \\
&&
	 + 186624 \hat{t}^{11}),
\\
a_{3}&=&-2 \hat{t} (57 \hat{s}^{10} + 1233 \hat{s}^9 \hat{t} 
       + 46541 \hat{s}^8 \hat{t}^2 + 513120 \hat{s}^7 \hat{t}^3 
       + 2646793 \hat{s}^6 \hat{t}^4 + 7942109 \hat{s}^5 \hat{t}^5
\nonumber \\
&&       + 15041136 \hat{s}^4 \hat{t}^6 + 18324922 \hat{s}^3 \hat{t}^7 
       + 13942380 \hat{s}^2 \hat{t}^8
       + 6013080 \hat{s} \hat{t}^9 + 1119744 \hat{t}^{10}),
\\
a_{4}&=& 2 (935 \hat{s}^{10} + 9398 \hat{s}^9 \hat{t} 
       + 117747 \hat{s}^8 \hat{t}^2 + 1103652 \hat{s}^7 \hat{t}^3 
       + 6182220 \hat{s}^6 \hat{t}^4 + 21423546 \hat{s}^5 \hat{t}^5
\nonumber \\
&&
       + 47491450 \hat{s}^4 \hat{t}^6 + 67574132 \hat{s}^3 \hat{t}^7 
       + 59508939 \hat{s}^2 \hat{t}^8 + 29339460 \hat{s} \hat{t}^9 
       + 6158916 \hat{t}^{10}),
\\
a_{5}&=&-2 (8039 \hat{s}^9 + 112887 \hat{s}^8 \hat{t} 
       + 1157014 \hat{s}^7 \hat{t}^2 + 7632256 \hat{s}^6 \hat{t}^3 
       + 31876569 \hat{s}^5 \hat{t}^4 
\nonumber \\
&&
       + 85147430 \hat{s}^4 \hat{t}^5
       + 144700858 \hat{s}^3 \hat{t}^6 + 150182520 \hat{s}^2 \hat{t}^7 
       + 85844772 \hat{s} \hat{t}^8 + 20531880 \hat{t}^9),
\\
a_{6}&=& 2 (43072 \hat{s}^8 + 638490 \hat{s}^7 \hat{t} 
       + 5393635 \hat{s}^6 \hat{t}^2 + 28486982 \hat{s}^5 \hat{t}^3 
       + 94986651 \hat{s}^4 \hat{t}^4 
\nonumber \\
&&
       + 198281780 \hat{s}^3 \hat{t}^5
       + 248119176 \hat{s}^2 \hat{t}^6 + 167349456 \hat{s} \hat{t}^7 
       + 46204020 \hat{t}^8),
\\
a_{7}&=&-2 (158802 \hat{s}^7 + 2143917 \hat{s}^6 \hat{t} 
       + 15477603 \hat{s}^5 \hat{t}^2 + 67698320 \hat{s}^4 \hat{t}^3 
       + 180289870 \hat{s}^3 \hat{t}^4 
\nonumber \\
&&
       + 280325328 \hat{s}^2 \hat{t}^5
       + 228221280 \hat{s} \hat{t}^6 + 73941984 \hat{t}^7),
\\
a_{8}&=& 775181 \hat{s}^6 + 9628777 \hat{s}^5 \hat{t} 
       + 60464369 \hat{s}^4 \hat{t}^2 + 217547464 \hat{s}^3 \hat{t}^3 
       + 438545220 \hat{s}^2 \hat{t}^4 
\nonumber \\
&&
       + 444319344 \hat{s} \hat{t}^5 + 172576656 \hat{t}^6,
\\
a_{9}&=&-2 (674202 \hat{s}^5 + 7783209 \hat{s}^4 \hat{t} 
       + 41993932 \hat{s}^3 \hat{t}^2 + 117212424 \hat{s}^2 \hat{t}^3 
       + 154359000 \hat{s} \hat{t}^4 
\nonumber \\
&&
       + 73984752 \hat{t}^5),
\\
a_{10}&=& 4 (446021 \hat{s}^4 + 4708219 \hat{s}^3 \hat{t} 
       + 20480415 \hat{s}^2 \hat{t}^2
       + 37508616 \hat{s} \hat{t}^3 + 23128740 \hat{t}^4),
\\
a_{11}&=&-8 (233734 \hat{s}^3 + 2111409 \hat{s}^2 \hat{t} 
       + 6071274 \hat{s} \hat{t}^2 + 5141880 \hat{t}^3),
\\
a_{12}&=& 6 (259913 \hat{s}^2 + 1570956 \hat{s} \hat{t} + 2057724 \hat{t}^2),
\\
a_{13}&=&-72 (11537 \hat{s} + 31194 \hat{t}),
\\
a_{14}&=& 187272.
\end{eqnarray*}
Note that the long-distance factor $\langle O_8^{H}({}^3S_1)\rangle^2$
does not appear in eq.~(\ref{eq:sub-2H}) because it has been factored
out in eq.~(\ref{xsec}).

\section{Parton-level cross sections
for $g g \to c \bar{c}_{n_1}({}^3S_1) + b \bar{b}_{n_2}({}^3S_1)$
\label{app:diff}}
In this appendix, we present the parton-level differential cross
sections for $pp\to H_1+H_2+X$ for $H_1=J/\psi$ and $H_2=\Upsilon$. 
In $pp\to J/\psi+\Upsilon+X$ we consider the three channels
$gg \to c\bar{c}_8({}^3S_1) + b\bar{b}_8({}^3S_1)$,
$gg \to c\bar{c}_1({}^3S_1) + b\bar{b}_8({}^3S_1)$, and
$gg \to c\bar{c}_8({}^3S_1) + b\bar{b}_1({}^3S_1)$.
The Mandelstam variables are defined in eq.~(\ref{eq:Mandelstam}),
where $p_1$ and $p_2$ are momenta for $c\bar{c}_{n_1}$ and $b\bar{b}_{n_2}$,
respectively.

\subsection{
$g g \to c \bar{c}_8({}^3S_1) + b \bar{b}_8 ({}^3S_1)$
\label{app:octoct}}
The parton-level differential cross section for 
$g g \to c \bar{c}_8({}^3S_1) + b \bar{b}_8 ({}^3S_1)$ is given by
\begin{equation}
\frac{d \hat{\sigma}}{d \hat{t}}
[g g \to c \bar{c}_8({}^3S_1) + b \bar{b}_8 ({}^3S_1)] =
F_1 \sum_{i,j=0}^{8} b_{ij} m_{J/\psi}^{2 i} m_{\Upsilon}^{2 j},
\end{equation}
where
\begin{equation}
F_1 =
\frac{\pi^3 \alpha_s^4 }
{108 m_{J/\psi}^3 m_{\Upsilon}^3 \hat{s}^2 
(\hat{t}-m_{J/\psi}^2)^2 
(\hat{t}-m_{\Upsilon}^2)^2 
(\hat{u}-m_{J/\psi}^2)^2 
(\hat{u}-m_{\Upsilon}^2)^2 
[\hat{s}^2-(m_{J/\psi}^2-m_{\Upsilon}^2)^2]^2}.
\end{equation}
Here, $b_{ij}=b_{ji}$  and non-vanishing elements of $b_{ij}$'s are
\begin{eqnarray*}
b_{0 0} &=& 54 \hat{s}^2 \hat{t}^2 (\hat{s}+\hat{t})^2 
(\hat{s}^2+\hat{s} \hat{t}+\hat{t}^2)^3,
\\
b_{0 1} &=& -54 \hat{s} \hat{t}^2 (\hat{s}+\hat{t}) 
(\hat{s}^2+\hat{s} \hat{t}+\hat{t}^2)
(3 \hat{s}^5+6 \hat{s}^4 \hat{t}+6 \hat{s}^3 \hat{t}^2
+2 \hat{s}^2 \hat{t}^3-\hat{s} \hat{t}^4-2 \hat{t}^5),
\\
b_{0 2} &=& 2 \hat{t}^2 (46 \hat{s}^8+49 \hat{s}^7 \hat{t}
-199 \hat{s}^6 \hat{t}^2-701 \hat{s}^5 \hat{t}^3 -1090 \hat{s}^4 \hat{t}^4
-1004 \hat{s}^3 \hat{t}^5-548 \hat{s}^2 \hat{t}^6-135 \hat{s} \hat{t}^7
\nonumber \\
&&
+27 \hat{t}^8),
\\
b_{0 3} &=& 2 \hat{t}^2 (97 \hat{s}^7+499 \hat{s}^6 \hat{t}
+1147 \hat{s}^5 \hat{t}^2+1520 \hat{s}^4 \hat{t}^3 +1218 \hat{s}^3 \hat{t}^4
+508 \hat{s}^2 \hat{t}^5-16 \hat{s} \hat{t}^6-135 \hat{t}^7),
\\
b_{0 4} &=& -2 \hat{t}^2 (154 \hat{s}^6+484 \hat{s}^5 \hat{t}
+636 \hat{s}^4 \hat{t}^2+323 \hat{s}^3 \hat{t}^3 -177 \hat{s}^2 \hat{t}^4
-410 \hat{s} \hat{t}^5-289 \hat{t}^6),
\\
b_{0 5} &=& 2 \hat{t}^2 (49 \hat{s}^5-26 \hat{s}^4 \hat{t}
-350 \hat{s}^3 \hat{t}^2-634 \hat{s}^2 \hat{t}^3 -613 \hat{s} \hat{t}^4
-346 \hat{t}^5),
\\
b_{0 6} &=& 2 \hat{t}^2 (62 \hat{s}^4+281 \hat{s}^3 \hat{t}
+465 \hat{s}^2 \hat{t}^2+430 \hat{s} \hat{t}^3 +249 \hat{t}^4),
\\
b_{0 7} &=& -2 \hat{t}^2 (65 \hat{s}^3+149 \hat{s}^2 \hat{t}
+149 \hat{s} \hat{t}^2+103 \hat{t}^3),
\\
b_{0 8} &=& 38 \hat{t}^2 (\hat{s}^2+\hat{s} \hat{t}+\hat{t}^2),
\\
b_{1 1} &=& \hat{t} (54 \hat{s}^9+724 \hat{s}^8 \hat{t}
+2167 \hat{s}^7 \hat{t}^2+3438 \hat{s}^6 \hat{t}^3+3305 \hat{s}^5 \hat{t}^4
+1685 \hat{s}^4 \hat{t}^5+68 \hat{s}^3 \hat{t}^6-361 \hat{s}^2 \hat{t}^7
\nonumber \\
&&
+216 \hat{t}^9),
\\
b_{1 2} &=& -\hat{t} (152 \hat{s}^8+1025 \hat{s}^7 \hat{t}
+2436 \hat{s}^6 \hat{t}^2+3632 \hat{s}^5 \hat{t}^3 +3862 \hat{s}^4 \hat{t}^4
+3159 \hat{s}^3 \hat{t}^5+2468 \hat{s}^2 \hat{t}^6
\nonumber \\
&&
+2074 \hat{s} \hat{t}^7 +1350 \hat{t}^8),
\\
b_{1 3} &=& \hat{t} (34 \hat{s}^7+212 \hat{s}^6 \hat{t}
+1089 \hat{s}^5 \hat{t}^2+3503 \hat{s}^4 \hat{t}^3 +6172 \hat{s}^3 \hat{t}^4
+7093 \hat{s}^2 \hat{t}^5+5980 \hat{s} \hat{t}^6+3439 \hat{t}^7),
\\
b_{1 4} &=& \hat{t} (200 \hat{s}^6+314 \hat{s}^5 \hat{t}
-1478 \hat{s}^4 \hat{t}^2-5353 \hat{s}^3 \hat{t}^3 -8018 \hat{s}^2 \hat{t}^4
-7667 \hat{s} \hat{t}^5-4728 \hat{t}^6),
\\
b_{1 5} &=& -\hat{t} (18 \hat{s}^5-740 \hat{s}^4 \hat{t}
-3227 \hat{s}^3 \hat{t}^2-5442 \hat{s}^2 \hat{t}^3 -5707 \hat{s} \hat{t}^4
-3962 \hat{t}^5),
\\
b_{1 6} &=& -\hat{t} (340 \hat{s}^4+1469 \hat{s}^3 \hat{t}
+2662 \hat{s}^2 \hat{t}^2+2833 \hat{s} \hat{t}^3+2194 \hat{t}^4),
\\
b_{1 7} &=& \hat{t} (298 \hat{s}^3+780 \hat{s}^2 \hat{t}+893 \hat{s} \hat{t}^2
+807 \hat{t}^3),
\\
b_{1 8} &=& -38 \hat{t} (2 \hat{s}^2+3 \hat{s} \hat{t}+4 \hat{t}^2),
\\
b_{2 2} &=& 152 \hat{s}^8+1243 \hat{s}^7 \hat{t}+5142 \hat{s}^6 \hat{t}^2
+12412 \hat{s}^5 \hat{t}^3+20633 \hat{s}^4 \hat{t}^4 +24264 \hat{s}^3 \hat{t}^5
+20866 \hat{s}^2 \hat{t}^6 
\nonumber \\
&&
+13616 \hat{s} \hat{t}^7 +6546 \hat{t}^8,
\\
b_{2 3} &=& -304 \hat{s}^7-2502 \hat{s}^6 \hat{t}-9334 \hat{s}^5 \hat{t}^2
-22006 \hat{s}^4 \hat{t}^3-34432 \hat{s}^3 \hat{t}^4 -36842 \hat{s}^2 \hat{t}^5
-27395 \hat{s} \hat{t}^6
\nonumber \\
&&
-14020 \hat{t}^7,
\\
b_{2 4} &=& 146 \hat{s}^6+1808 \hat{s}^5 \hat{t}+7883 \hat{s}^4 \hat{t}^2
+19046 \hat{s}^3 \hat{t}^3 +27583 \hat{s}^2 \hat{t}^4+25828 \hat{s} \hat{t}^5
+15958 \hat{t}^6,
\\
b_{2 5} &=& -2 (40 \hat{s}^5+380 \hat{s}^4 \hat{t}+2229 \hat{s}^3 \hat{t}^2
+4889 \hat{s}^2 \hat{t}^3 +6304 \hat{s} \hat{t}^4+5163 \hat{t}^5),
\\
b_{2 6} &=& 216 \hat{s}^4+915 \hat{s}^3 \hat{t}+2363 \hat{s}^2 \hat{t}^2
+3680 \hat{s} \hat{t}^3+4078 \hat{t}^4,
\\
b_{2 7} &=& -168 \hat{s}^3-666 \hat{s}^2 \hat{t}-891 \hat{s} \hat{t}^2
-1168 \hat{t}^3,
\\
b_{2 8} &=& 38 (\hat{s}^2+3 \hat{s} \hat{t}+6 \hat{t}^2),
\\
b_{3 3} &=& 1324 \hat{s}^6+7637 \hat{s}^5 \hat{t}+23669 \hat{s}^4 \hat{t}^2
+45278 \hat{s}^3 \hat{t}^3 +57940 \hat{s}^2 \hat{t}^4+48610 \hat{s} \hat{t}^5
+27204 \hat{t}^6,
\\
b_{3 4} &=& -1306 \hat{s}^5-8050 \hat{s}^4 \hat{t}-22644 \hat{s}^3 \hat{t}^2
-38898 \hat{s}^2 \hat{t}^3 -40961 \hat{s} \hat{t}^4-28098 \hat{t}^5,
\\
b_{3 5} &=& 148 \hat{s}^4+3198 \hat{s}^3 \hat{t}+9967 \hat{s}^2 \hat{t}^2
+16183 \hat{s} \hat{t}^3+15507 \hat{t}^4,
\\
b_{3 6} &=& -2 (4 \hat{s}^3+230 \hat{s}^2 \hat{t}+1321 \hat{s} \hat{t}^2
+2166 \hat{t}^3),
\\
b_{3 7} &=& 184 \hat{s}^2+295 \hat{s} \hat{t}+722 \hat{t}^2,
\\
b_{3 8} &=& -38 (\hat{s}+4 \hat{t}),
\\
b_{4 4} &=& 2651 \hat{s}^4+10864 \hat{s}^3 \hat{t}+25178 \hat{s}^2 \hat{t}^2
+32284 \hat{s} \hat{t}^3+27180 \hat{t}^4,
\\
b_{4 5} &=& -1267 \hat{s}^3-6028 \hat{s}^2 \hat{t}-11581 \hat{s} \hat{t}^2
-13788 \hat{t}^3,
\\
b_{4 6} &=& -171 \hat{s}^2+1276 \hat{s} \hat{t}+2998 \hat{t}^2,
\\
b_{4 7} &=& \hat{s}-138 \hat{t},
\\
b_{4 8} &=& 38,
\\
b_{5 5} &=& 1665 \hat{s}^2+3866 \hat{s} \hat{t}+6684 \hat{t}^2,
\\
b_{5 6} &=& -341 \hat{s}-1330 \hat{t},
\\
b_{5 7} &=& -17,
\\
b_{6 6} &=& 282.
\end{eqnarray*}

\subsection{
$g g \to c \bar{c}_1({}^3S_1) + b \bar{b}_8 ({}^3S_1)$
\label{app:singoct18}}
The parton-level differential cross section for 
$g g \to c \bar{c}_1({}^3S_1) + b \bar{b}_8 ({}^3S_1)$ is given by
\begin{equation}
\frac{d \hat{\sigma}}{d \hat{t}}
[g g \to c \bar{c}_1({}^3S_1) + b \bar{b}_8 ({}^3S_1)] =
F_2 \sum_{i,j=0}^{3} c_{ij} m_{J/\psi}^{2 i} m_{\Upsilon}^{2 j},
\end{equation}
where
\begin{equation}
F_2 =
\frac{10 \pi^3 \alpha_s^4 }
{243 m_{J/\psi} m_{\Upsilon}^3 \hat{s}^2 
(\hat{t}-m_{J/\psi}^2)^2 
(\hat{u}-m_{J/\psi}^2)^2 
(\hat{s}-m_{J/\psi}^2+m_{\Upsilon}^2)^2},
\end{equation}
and non-vanishing elements of $c_{ij}$'s are
\begin{subequations}
\label{eq:cij}
\begin{eqnarray}
c_{0 1} &=& \hat{t}^2 (\hat{s}+\hat{t})^2,
\\
c_{0 2} &=& -2 \hat{t} (\hat{s}+\hat{t})^2,
\\
c_{0 3} &=& \hat{t} (2 \hat{s}+\hat{t}),
\\
c_{1 0} &=& 2 (\hat{s}^2+\hat{s} \hat{t}+\hat{t}^2)^2,
\\
c_{1 1} &=& -2 \hat{t}^2 (\hat{s}+3 \hat{t}),
\\
c_{1 2} &=& 2 (\hat{s}^2+3 \hat{t}^2),
\\
c_{1 3} &=& -2 (\hat{s}+\hat{t}),
\\
c_{2 0} &=& -2 (\hat{s}+\hat{t}) (2 \hat{s}^2+\hat{s} \hat{t}+2 \hat{t}^2),
\\
c_{2 1} &=& 3 \hat{s}^2+2 \hat{s} \hat{t}+9 \hat{t}^2,
\\
c_{2 2} &=& 2 (2 \hat{s}-3 \hat{t}),
\\
c_{2 3} &=& 1,
\\
c_{3 0} &=& 2 (\hat{s}^2+\hat{s} \hat{t}+\hat{t}^2),
\\
c_{3 1} &=& -2 (\hat{s}+2 \hat{t}),
\\
c_{3 2} &=& 2.
\end{eqnarray}
\end{subequations}
\subsection{
$g g \to c \bar{c}_8({}^3S_1) + b \bar{b}_1 ({}^3S_1)$
\label{app:singoct81}}
The parton-level differential cross section for 
$g g \to c \bar{c}_8({}^3S_1) + b \bar{b}_1 ({}^3S_1)$ is given by
\begin{equation}
\frac{d \hat{\sigma}}{d \hat{t}}
[g g \to c \bar{c}_8({}^3S_1) + b \bar{b}_1 ({}^3S_1)] =
F_3 \sum_{i,j=0}^{3} c_{ji} m_{J/\psi}^{2 i} m_{\Upsilon}^{2 j},
\end{equation}
where $c_{ji}$'s are defined in eq.~(\ref{eq:cij})  and
\begin{equation}
F_3 =
\frac{10 \pi^3 \alpha_s^4 }
{243 m_{J/\psi}^3 m_{\Upsilon} \hat{s}^2 
(\hat{t}-m_{\Upsilon}^2)^2 
(\hat{u}-m_{\Upsilon}^2)^2 
(\hat{s}+m_{J/\psi}^2-m_{\Upsilon}^2)^2}.
\end{equation}


\begin{thebibliography}{999}
\bibitem{Einhorn:1975ua}
  M.B.~Einhorn and S.D.~Ellis,
  {\it Hadronic production of the new resonances: probing gluon distributions},
  {\it Phys.\ Rev.\ }  {\bf D 12} (1975) 2007.

\bibitem{Ellis:1976fj}
  S.D.~Ellis, M.B.~Einhorn and C.~Quigg,
  {\it Comment on hadronic production of psions},
  {\it Phys.\ Rev.\ Lett.\ }  {\bf 36} (1976) 1263.

\bibitem{Chang:1979nn}
  C.-H.~Chang,
  {\it Hadronic production of $J/\psi$ associated with a gluon},
  {\it Nucl.\ Phys.\ }  {\bf B 172} (1980) 425.

\bibitem{Berger:1980ni}
  E.L.~Berger and D.L.~Jones,
  {\it Inelastic photoproduction of $J/\psi$ and $\Upsilon$ by gluons},
  {\it Phys.\ Rev.\ }  {\bf D 23} (1981) 1521.

\bibitem{Baier:1981zz}
  R.~Baier and R.~Ruckl,
  {\it On inelastic leptoproduction of heavy quarkonium states},
  {\it Nucl.\ Phys.\ } {\bf B 201} (1982) 1.

\bibitem{Bodwin:1994jh}
  G.T.~Bodwin, E.~Braaten and G.P.~Lepage,
  {\it Rigorous QCD analysis of inclusive annihilation and production of heavy
   quarkonium},
  {\it Phys.\ Rev.\ } {\bf D 51} (1995) 1125
  [Erratum-ibid.\  {\bf D 55} (1997) 5853]
  [hep-ph/9407339].

\bibitem{Fritzsch:1977ay}
  H.~Fritzsch,
  {\it Producing heavy quark flavors in hadronic collisions: a test of quantum
  chromodynamics},
  {\it Phys.\ Lett.\ } {\bf B 67} (1977) 217.

\bibitem{Halzen:1977rs}
  F.~Halzen,
  {\it CVC for gluons and hadroproduction of quark flavors},
  {\it Phys.\ Lett.\ } {\bf B 69} (1977) 105.

\bibitem{Gluck:1977zm}
  M.~Gluck, J.F.~Owens and E.~Reya,
  {\it Gluon contribution to hadronic $J/\psi$ production},
 {\it Phys.\ Rev.\ } {\bf D 17} (1978) 2324.

\bibitem{Barger:1979js}
  V.D.~Barger, W.-Y.~Keung and R.J.N.~Phillips,
  {\it On $\psi$ and $\Upsilon$ production via gluons},
  {\it Phys.\ Lett.\ } {\bf B 91} (1980) 253.

\bibitem{Bodwin:1992ye}
  G.T.~Bodwin, E.~Braaten and G.P.~Lepage,
  {\it Rigorous QCD predictions for decays of P wave quarkonia},
  {\it Phys.\ Rev.\ } {\bf D 46} (1992) 1914
  [hep-lat/9205006].

\bibitem{Braaten:1994vv}
  E.~Braaten and S.~Fleming,
  {\it Color octet fragmentation and the $\psi^\prime$ surplus at the Tevatron},
  {\it Phys.\ Rev.\ Lett.\ } {\bf 74} (1995) 3327
  [hep-ph/9411365].
\bibitem{Braaten:1999qk}
  E.~Braaten, B.A.~Kniehl and J.~Lee,
  {\it Polarization of prompt $J/\psi$ at the Tevatron},
  {\it Phys.\ Rev.\ } {\bf D 62} (2000) 094005
  [hep-ph/9911436].

\bibitem{Affolder:2000nn}
  CDF collaboration, A.A.~Affolder et al.,
  {\it Measurement of $J/\psi$ and $\psi(2S)$ polarization in $p\bar{p}$
   collisions at $\sqrt{s} =$ \rm{1.8} TeV},
  {\it Phys.\ Rev.\ Lett.\ }  {\bf 85} (2000) 2886
  [hep-ex/0004027].
\bibitem{Abulencia:2007us}
  CDF collaboration, A.~Abulencia et al.,
  {\it Polarization of $J/\psi$ and $\psi(2S)$ mesons produced in $p \bar{p}$
  collisions at $\sqrt{s}$ = \rm{1.96}\,TeV},
  {\it Phys.\ Rev.\ Lett.\ }  {\bf 99} (2007) 132001
  [arXiv:0704.0638].
\bibitem{review} 
  Quarkonium Working Group collaboration, N.~Brambilla et al.,
  {\it Heavy quarkonium physics},
  hep-ph/0412158.

\bibitem{Campbell:2007ws}
  J.M.~Campbell, F.~Maltoni and F.~Tramontano,
  {\it QCD corrections to $J/\psi$ and $\Upsilon$ production 
  at hadron colliders},
  {\it Phys.\ Rev.\ Lett.\ }  {\bf 98} (2007) 252002
  [hep-ph/0703113].
\bibitem{Gong:2008sn}
  B.~Gong and J.-X.~Wang,
 {\it Next-to-leading-order QCD corrections to $J/\psi$ polarization at Tevatron
  and Large-Hadron-Collider energies},
  {\it Phys.\ Rev.\ Lett.\ } {\bf 100} (2008) 232001
  [arXiv:0802.3727].

\bibitem{Gong:2008hk}
 B.~Gong and J.-X.~Wang,
 {\it QCD corrections to polarization of $J/\psi$ and $\Upsilon$ at Tevatron and
  LHC},
  {\it Phys.\ Rev.\ } {\bf D 78} (2008) 074011
  [arXiv:0805.2469].

\bibitem{Gong:2008ft}
  B.~Gong, X.Q.~Li and J.-X.~Wang,
  {\it QCD corrections to $J/\psi$ production via color octet states at Tevatron
  and LHC},
  {\it Phys.\ Lett.\ } {\bf B 673} (2009) 197
  [Erratum-ibid.\  {\bf B 693} (2010) 612]
  [arXiv:0805.4751].

\bibitem{Ma:2008gq}
  Y.-Q.~Ma, Y.-J.~Zhang and K.-T.~Chao,
  {\it QCD correction to $e^+ e^- \to J/\psi g g$ at $B$ Factories},
  {\it Phys.\ Rev.\ Lett.\ } {\bf 102} (2009) 162002
  [arXiv:0812.5106].

\bibitem{Gong:2009kp}
  B.~Gong and J.-X.~Wang,
  {\it Next-to-leading-order QCD corrections to $e^ + e^- \to J/\psi+gg$ at 
  the $B$ Factories},
  {\it Phys.\ Rev.\ Lett.\ } {\bf 102} (2009) 162003
  [arXiv:0901.0117].

\bibitem{:2009nj}
  BELLE collaboration, P.~Pakhlov et al.,
  {\it Measurement of the $e^+e^- \to J/\psi c\bar{c}$ cross section 
  at $\sqrt{s} \approx$ 10.6 GeV},
  {\it Phys.\ Rev.\ } {\bf D 79} (2009) 071101
  [arXiv:0901.2775].

\bibitem{Zhang:2009ym}
  Y.-J.~Zhang, Y.-Q.~Ma, K.~Wang and K.-T.~Chao,
 {\it QCD radiative correction to color-octet $J/\psi$ inclusive production 
 at $B$ Factories},
  {\it Phys.\ Rev.\ } {\bf D 81} (2010) 034015
  [arXiv:0911.2166].

\bibitem{Artoisenet:2009xh}
  P.~Artoisenet, J.M.~Campbell, F.~Maltoni and F.~Tramontano,
  {\it $J/\psi$ production at HERA},
  {\it Phys.\ Rev.\ Lett.\ } {\bf 102} (2009) 142001
  [arXiv:0901.4352].

\bibitem{Chang:2009uj}
  C.-H.~Chang, R.~Li and J.-X.~Wang,
 {\it $J/\psi$ polarization in photo-production up-to the next-to-leading 
  order of QCD},
  {\it Phys.\ Rev.\ } {\bf D 80} (2009) 034020
  [arXiv:0901.4749].

\bibitem{Butenschoen:2009zy}
  M.~Butenschoen and B.A.~Kniehl,
  {\it Complete next-to-leading-order corrections to $J/\psi$ photoproduction in
  nonrelativistic quantum chromodynamics},
  {\it Phys.\ Rev.\ Lett.\ } {\bf 104} (2010) 072001
  [arXiv:0909.2798].

\bibitem{Abe:2002rb}
  BELLE collaboration, K.~Abe et al.,
  {\it Observation of double $c\bar{c}$ production in $e^+ e^-$ annihilation
  at $\sqrt{s}\approx$ 10.6\, GeV},
  {\it Phys.\ Rev.\ Lett.\ } {\bf 89} (2002) 142001
  [hep-ex/0205104].

\bibitem{double1}
  E.~Braaten and J.~Lee,
  {\it Exclusive double-charmonium production in $e^+ e^-$ annihilation},
  {\it Phys.\ Rev.\ } {\bf D 67} (2003) 054007
  [Erratum-ibid.\ {\bf D 72} (2005) 099901]
  [hep-ph/0211085].

\bibitem{Liu:2002wq}
  K.-Y.~Liu, Z.-G.~He and K.-T.~Chao,
  {\it Problems of double charm production in $e^+ e^-$ annihilation at 
  $\sqrt{s}=$ \rm{10.6}\,GeV},
  {\it Phys.\ Lett.\ } {\bf B 557} (2003) 45
  [hep-ph/0211181].

\bibitem{Hagiwara:2003cw}
  K.~Hagiwara, E.~Kou and C.-F.~Qiao,
  {\it Exclusive $J/\psi$ productions at $e^{+} e^{-}$ colliders},
  {\it Phys.\ Lett.\ } {\bf B 570} (2003) 39
  [hep-ph/0305102].

\bibitem{Brodsky:2003hv}
  S.J.~Brodsky, A.S.~Goldhaber and J.~Lee,
  {\it Hunting for glueballs in electron positron annihilation},
  {\it Phys.\ Rev.\ Lett.\ } {\bf 91} (2003) 112001
  [hep-ph/0305269].

\bibitem{Bodwin:2002fk}
  G.T.~Bodwin, J.~Lee and E.~Braaten,
  {\it $e^+ e^-$ annihilation into $J/\psi + J/\psi$},
  {\it Phys.\ Rev.\ Lett.\ } {\bf 90} (2003) 162001
  [hep-ph/0212181].

\bibitem{Bodwin:2002kk}
  G.T.~Bodwin, J.~Lee and E.~Braaten,
  {\it Exclusive double-charmonium production from $e^+ e^-$
   annihilation into two virtual photons},
  {\it Phys.\ Rev.\ } {\bf D 67} (2003) 054023
  [Erratum-ibid.\  {\bf D 72} (2005) 099904]
  [hep-ph/0212352].

\bibitem{Bodwin:2006yd}
  G.T.~Bodwin, E.~Braaten, J.~Lee and C.~Yu,
  {\it Exclusive two-vector-meson production from $e^+ e^-$ annihilation},
  {\it Phys.\ Rev.\ } {\bf D 74} (2006) 074014
  [hep-ph/0608200].

\bibitem{Ma:2004qf}
  J.P.~Ma and Z.G.~Si,
{\it Predictions for $e^+e^- \to J/\psi \eta_c$ with light-cone wave-functions},
  {\it Phys.\ Rev.\ } {\bf D 70} (2004) 074007
  [hep-ph/0405111].

\bibitem{Bondar:2004sv}
  A.E.~Bondar and V.L.~Chernyak,
  {\it Is the BELLE result for the cross section 
  $\sigma(e^+ e^- \to J/\psi +  \eta_c)$ a
  real difficulty for QCD?},
  {\it Phys.\ Lett.\ } {\bf B 612} (2005) 215
  [hep-ph/0412335].

\bibitem{Braguta:2008tg}
  V.V.~Braguta,
  {\it Double charmonium production at $B$-factories 
   within light cone formalism},
  {\it Phys.\ Rev.\ } {\bf D 79} (2009) 074018
  [arXiv:0811.2640].

\bibitem{Zhang:2005cha}
  Y.-J.~Zhang, Y.-j.~Gao and K.-T.~Chao,
  {\it Next-to-leading order QCD correction to $e^+ e^- \to J/\psi + \eta_c$ at
  $\sqrt{s}=$ \rm{10.6}\,GeV},
  {\it Phys.\ Rev.\ Lett.\ } {\bf 96} (2006) 092001
  [hep-ph/0506076].

\bibitem{Gong:2008ce}
  B.~Gong and J.-X.~Wang,
{\it QCD corrections to double $J/\psi$ production in $e^{+} e^{-}$ annihilation
  at $\sqrt{s}$ = \rm{10.6} GeV},
  {\it Phys.\ Rev.\ Lett.\ } {\bf 100} (2008) 181803
  [arXiv:0801.0648].

\bibitem{He:2007te}
  Z.-G.~He, Y.~Fan and K.-T.~Chao,
  {\it Relativistic corrections to $J/\psi$ exclusive and inclusive
  double charm production at $B$ factories},
  {\it Phys.\ Rev.\ } {\bf D 75} (2007) 074011
  [hep-ph/0702239].

\bibitem{Bodwin:2007ga}
  G.T.~Bodwin, J.~Lee and C.~Yu,
  {\it Resummation of relativistic corrections to $e^+ e^- \to J/\psi+\eta_c$},
  {\it Phys.\ Rev.\ } {\bf D 77} (2008) 094018
  [arXiv:0710.0995].

\bibitem{Barger:1995vx}
  V.D.~Barger, S.~Fleming and R.J.N.~Phillips,
  {\it Double gluon fragmentation to $J/\psi$ pairs at the Tevatron},
  {\it Phys.\ Lett.\ } {\bf B 371} (1996) 111
  [hep-ph/9510457].

\bibitem{Qiao:2002rh}
  C.-F.~Qiao,
  {\it $J/\psi$ pair production at the Tevatron},
  {\it Phys.\ Rev.\ } {\bf D 66} (2002) 057504
  [hep-ph/0206093].
  
\bibitem{Li:2009ug}
  R.~Li, Y.-J.~Zhang and K.-T.~Chao,
  {\it Pair production of heavy quarkonium and $B_c^{(\ast)}$ mesons at hadron
  colliders},
  {\it Phys.\ Rev.\ } {\bf D 80} (2009) 014020
  [arXiv:0903.2250].

\bibitem{Qiao:2009kg}
  C.-F.~Qiao, L.-P.~Sun and P.~Sun,
  {\it Testing charmonium production mechanism via polarized $J/\psi$ pair
  production at the LHC},
  {\it J.\ Phys.\ } {\bf G 37} (2010) 075019
  [arXiv:0903.0954].

\bibitem{Ko:2010vh}
  P.~Ko, C.~Yu and J.~Lee,
  {\it $p p \to J/\psi+\Upsilon+X$ as a clean probe to the quarkonium production
  mechanism},
  arXiv:1006.3846.

\bibitem{Collins:1981uk}
  J.C.~Collins and D.E.~Soper,
  {\it Back-to-back jets in QCD},
  {\it Nucl.\ Phys.\ } {\bf B 193} (1981) 381
  [Erratum-ibid.\  {\bf B 213} (1983) 545].
 
\bibitem{Braaten:2000pc}
  E.~Braaten and J.~Lee,
  {\it Next-to-leading order calculation of the color octet ${}^3S_1$ gluon
  fragmentation function for heavy quarkonium},
  {\it Nucl.\ Phys.\ } {\bf B 586} (2000) 427
  [hep-ph/0004228].

\bibitem{Lee:2005jw}
  J.~Lee,
  {\it Next-to-leading order calculation of a fragmentation function in a
  light-cone gauge},
  {\it Phys.\ Rev.\ } {\bf D 71} (2005) 094007
  [hep-ph/0504285].

\bibitem{Gribov:1972ri}
  V.N.~Gribov and L.N.~Lipatov,
  {\it Deep inelastic $ep$ scattering in perturbation theory},
  {\it Sov.\ J.\ Nucl.\ Phys.\ } {\bf 15} (1972) 438
  [Yad.\ Fiz.\  {\bf 15} (1972) 781].

\bibitem{Altarelli:1977zs}
  G.~Altarelli and G.~Parisi,
  {\it Asymptotic freedom in parton language},
  {\it Nucl.\ Phys.\ } {\bf B 126} (1977) 298.

\bibitem{Dokshitzer:1977sg}
  Y.L.~Dokshitzer,
  {\it Calculation of the structure functions for deep inelastic scattering and 
    $e^+e^-$ annihilation by perturbation theory in quantum chromodynamics},
  {\it Sov.\ Phys.\ JETP } {\bf 46} (1977) 641
  [Zh.\ Eksp.\ Teor.\ Fiz.\  {\bf 73} (1977) 1216].

\bibitem{Bodwin:2006dn}
  G.T.~Bodwin, D.~Kang and J.~Lee,
  {\it Potential-model calculation of an order-$v^2$ NRQCD matrix element},
  {\it Phys.\ Rev.\ } {\bf D 74} (2006) 014014
  [hep-ph/0603186].

\bibitem{Bodwin:2006dm}
  G.T.~Bodwin, D.~Kang and J.~Lee,
  {\it Reconciling the light-cone and NRQCD approaches to calculating 
   $e^+ e^- \to J/\psi + \eta_c$},
  {\it Phys.\ Rev.\ } {\bf D 74} (2006) 114028
  [hep-ph/0603185].
 
\bibitem{Bodwin:2007fz}
  G.T.~Bodwin, H.S.~Chung, D.~Kang, J.~Lee and C.~Yu,
  {\it Improved determination of color-singlet nonrelativistic QCD matrix
  elements for $S$-wave charmonium},
  {\it Phys.\ Rev.\ } {\bf D 77} (2008) 094017
  [arXiv:0710.0994].

\bibitem{Bodwin:2008vp}
  G.T.~Bodwin, H.S.~Chung, J.~Lee and C.~Yu,
  {\it Order-$\alpha_s$ corrections to the quarkonium electromagnetic
  current at all orders in the heavy-quark velocity},
  {\it Phys.\ Rev.\ } {\bf D 79} (2009) 014007
  [arXiv:0807.2634].

\bibitem{Chung:2008sm}
  H.S.~Chung, J.~Lee and D.~Kang,
  {\it Cornell potential parameters for $S$-wave heavy quarkonia},
  {\it J.\ Korean Phys.\ Soc.\ } {\bf 52} (2008) 1151
  [arXiv:0803.3116].
  

\bibitem{Kang:2007uv}
  D.~Kang, T.~Kim, J.~Lee and C.~Yu,
  {\it Inclusive charm production in $\Upsilon(nS)$ decay},
  {\it Phys.\ Rev.\ } {\bf D 76} (2007) 114018
  [arXiv:0707.4056].

\bibitem{Chung:2008yf}
  H.S.~Chung, T.~Kim and J.~Lee,
  {\it Invariant-mass distribution of $c\bar{c}$ in 
  $\Upsilon(1S) \to  c\bar{c}+ X$},
  {\it Phys.\ Rev.\ } {\bf D 78} (2008) 114027
  [arXiv:0805.1989].

\bibitem{Braaten:2000cm}
  E.~Braaten, S.~Fleming and A.K.~Leibovich,
  {\it NRQCD analysis of bottomonium production at the Tevatron},
  {\it Phys.\ Rev.\ } {\bf D 63} (2001) 094006
  [hep-ph/0008091].
\bibitem{Braaten:2000gw}
  E.~Braaten and J.~Lee,
  {\it Polarization of $\Upsilon(nS)$ at the Tevatron},
  {\it Phys.\ Rev.\ } {\bf D 63} (2001) 071501
  [hep-ph/0012244].
\bibitem{Kramer:2001hh}
  M.~Kr\"amer,
  {\it Quarkonium production at high-energy colliders},
  {\it Prog.\ Part.\ Nucl.\ Phys.\ } {\bf 47} (2001) 141
  [hep-ph/0106120].

\bibitem{Pumplin:2002vw}
  J.~Pumplin, et al.,
  {\it New generation of parton distributions with uncertainties from global QCD
   analysis},
  {\it JHEP } {\bf 07} (2002) 012
  [hep-ph/0201195].

\bibitem{Bodwin:2002hg}
  G.T.~Bodwin and A.~Petrelli,
  {\it Order-$v^4$ corrections to $S$-wave quarkonium decay},
  {\it Phys.\ Rev.\ } {\bf D 66} (2002) 094011
  [hep-ph/0205210].

\bibitem{Bodwin:2003wh}
  G.T.~Bodwin and J.~Lee,
  {\it Relativistic corrections to gluon fragmentation into spin-triplet
   $S$-wave quarkonium},
  {\it Phys.\ Rev.\ } {\bf D 69} (2004) 054003
  [hep-ph/0308016].

\bibitem{Bodwin:2009cb}
  G.T.~Bodwin, X.~Garcia i Tormo and J.~Lee,
  {\it Factorization of low-energy gluons in exclusive processes},
  {\it Phys.\ Rev.\ } {\bf D 81} (2010) 114005
  [arXiv:0903.0569].

\bibitem{Bodwin:2010fi}
  G.T.~Bodwin, X.~Garcia i Tormo and J.~Lee,
  {\it Factorization in exclusive quarkonium production},
  {\it Phys.\ Rev.\ } {\bf D 81} (2010) 114014
  [arXiv:1003.0061].
   
\bibitem{Artoisenet:2007xi}
  P.~Artoisenet, J.P.~Lansberg and F.~Maltoni,
  {\it Hadroproduction of $J/\psi$ and $\Upsilon$ in association with a
  heavy-quark pair},
  {\it Phys.\ Lett.\ } {\bf B 653} (2007) 60
  [hep-ph/0703129].

\bibitem{Amsler:2008zzb}
  Particle Data Group collaboration, C.~Amsler et al.,
  {\it Review of particle physics},
  {\it Phys.\ Lett.\ } {\bf B 667} (2008) 1.
  
\bibitem{Artoisenet:2008fc}
  P.~Artoisenet, J.M.~Campbell, J.P.~Lansberg, F.~Maltoni and F.~Tramontano,
  {\it $\Upsilon$ production at Fermilab Tevatron and LHC energies},
  {\it Phys.\ Rev.\ Lett.\ } {\bf 101} (2008) 152001
  [arXiv:0806.3282].


\end{thebibliography}
\end{document}